\documentclass[aps,prb,twocolumn,floats,showpacs]{revtex4}
\usepackage{epsfig}
\usepackage{amsmath}
\usepackage{amssymb}
\begin{document}
\title{Kondo effect in systems with dynamical symmetries}
\author{ T. Kuzmenko$^1$, K. Kikoin$^1$ and Y. Avishai$^{1,2}$\\ }
\affiliation {$^1$Department of Physics and $^2$Ilse Katz Center,
Ben-Gurion University, Beer-Sheva, Israel }
\date{\today}
\begin{abstract}
This paper is devoted to a systematic exposure of the Kondo
physics in quantum dots for which the low energy spin excitations
consist of a few different spin multiplets $|S_{i}M_{i} \rangle$.
Under certain conditions (to be explained below) some of the
lowest energy levels $E_{S_{i}}$ are nearly degenerate. The dot in
its ground state cannot then be regarded as a simple quantum top
in the sense that beside its spin operator other dot (vector)
operators ${\bf R}_{n}$ are needed (in order to fully determine
its quantum states), which have non-zero matrix elements between
states of different spin multiplets $\langle S_{i}M_{i} |{\bf
R}_{n}|S_{j}M_{j} \rangle \ne 0$. These "Runge-Lenz" operators do
not appear in the isolated dot-Hamiltonian (so in some sense they
are "hidden"). Yet, they are exposed when tunneling between dot
and leads is switched on. The effective spin Hamiltonian which
couples the metallic electron spin ${\bf s}$ with the operators of
the dot then contains new exchange terms, $J_{n} {\bf s} \cdot
{\bf R}_{n}$ beside the ubiquitous ones $J_{i} {\bf s}\cdot {\bf
S}_{i}$. The operators ${\bf S}_{i}$ and ${\bf R}_{n}$ generate a
dynamical group (usually $SO(n)$). Remarkably, the value of $n$
can be controlled by gate voltages, indicating that abstract
concepts such as dynamical symmetry groups are experimentally
realizable. Moreover, when an external magnetic field is applied
then, under favorable circumstances, the exchange interaction
involves solely the Runge-Lenz operators ${\bf R}_{n}$ and the
corresponding dynamical symmetry group is $SU(n)$. For example,
the celebrated group $SU(3)$ is realized in triple quantum dot
with four electrons.
\end{abstract}
\maketitle

\section{Introduction}

Recently, studies of the physical properties of artificially
fabricated nano-objects turn out to be a rapidly developing branch
of fundamental and applied physics. Progress in these fields is
stimulated both by the achievements of nanotechnology and by the
ambitious projects of information processing, data storage,
molecular electronics and spintronics. The corresponding
technological evolution enabled the fabrication of various
low-dimensional systems from semiconductor heterostructures to
quantum wires and constrictions, quantum dots (QD), molecular
bridges and artificial structures with large molecules built in
electric circuits \cite{Wire}. This impressive experimental
progress led to the development of nanophysics, a new aspect and
research direction in quantum physics \cite{Nano}. Artificial
nano-objects possess the familiar features of quantum mechanical
systems, but sometimes one may create in artificially fabricated
systems such conditions, which are hardly observable "in natura".
For example, $1D \to 2D$ crossover may be realized in quantum
networks \cite{Net} and constrictions \cite{Wees}. The Kondo
effect may be observed in non-equilibrium conditions \cite{Noneq},
at high magnetic fields \cite{Magn}, and at finite frequencies
\cite{Freq}. Moreover, a quantum dot in the Kondo regime can be
integrated into a circuit exhibiting the Aharonov-Bohm effect
\cite{Moti}.

In this paper we focus on an intriguing challenge  in this context
related to the specific symmetry of the nano-objects under study.
More precisely, one is interested in answering questions
pertaining to the nature of the underlying symmetry of the dot
Hamiltonian and the algebra of operators appearing in the exchange
Hamiltonian. The investigation of this topic is intimately related
with the geometric structure and electron occupation of the
quantum dot in its ground state. We refer to a quantum dot
composed of a single well and containing an odd number $N$ of
electrons as a {\it simple} quantum dot (SQD). The dot Hamiltonian
of a  SQD ( in the absence of an external magnetic field) is
composed of two degenerate levels and has an $SU(2)$ symmetry. In
that sense the symmetry is referred to as geometrical. The
exchange Hamiltonian is expressed in terms of the generators of
the group $SU(2)$. On the other hand, quantum dots containing a
single well with even $N$ or quantum dots containing several wells
are referred to as {\it composite} quantum dots (CQD). The low
energy states of an isolated CQD Hamiltonian are spin multiplets.
In the generic case, the only degeneracy is that of magnetic
quantum numbers. Yet, as we argue below, dot-lead tunneling
results in level renormalization and the emergence of an
additional degeneracy, both generic and accidental. To be more
precise, we note that 1) The exchange part of the Hamiltonian
includes the generators of a non-compact Lie group (usually
$SO(n)$ or $SU(n)$) and 2) The {\it renormalized} low-energy
spin-excitation levels of the CQD Hamiltonian are almost
degenerate (within a Kondo energy scale). These two aspects are
gathered under the term {\it dynamical symmetry}. A more
quantitative exposition will be presented below.

Experimentally, resonance Kondo tunneling was observed in QD with
odd electron occupation number under strong Coulomb blockade
\cite{KKK}, and in individual atoms and molecules deposited on
metallic surfaces and on the edges of metallic wires in
break-junction geometry \cite{Park}. According to the theory of
Kondo effect  in QD \cite{Glaz}, spin degrees of QD are involved
in Kondo resonance. In our notation these are SQD and the physics
of Kondo tunneling in this case is similar to that of Kondo
scattering in magnetically doped metals, at least in the regime of
linear response.

The Kondo physics seems to be richer in systems involving
tunneling through CQD. Our main purpose is to demonstrate that CQD
possesses dynamical symmetries whose realization in Kondo
tunneling is experimentally tangible. Such experimental tuning of
dynamical symmetries is not possible in conventional Kondo
scattering. In many cases even the very existence of Kondo
tunneling crucially depends on the dynamical spin symmetry of CQD.
Several models dealing with Kondo tunneling in CQDs possessing
dynamical symmetry were considered in our previous publications.
The case of $SO(4)$ symmetry in double quantum dot (DQD) was
studied in Refs. \onlinecite{KA01,KA02}. A more complicated case
of $SO(n)$ symmetry with variable $n$ in a triple quantum dot
(TQD) (composed of three potential valleys with intra-well
interaction and inter-well tunneling) is introduced in Ref.
\onlinecite{KKA02}.

The main goal here is to develop the general approach to the
problem of dynamical symmetries in Kondo tunneling through CQDs
and illustrate it by numerous examples of TQD in various
configurations, both in parallel and in series geometries. Within
this framework, our earlier results \cite{KKA02} fit into an
elegant pattern of classification of dynamical symmetry groups,
which is expanded here in a somewhat
more complete and rigorous formalism.
The main lesson to
be learned is that Kondo physics in CQD suggests a novel and in
some sense rather appealing aspect of low-dimensional physics of
interacting electrons. It substantiates, in a systematic way, that
dynamical symmetry groups play an important role in mesoscopic
physics. In particular, we encounter here some "famous" groups
which appear in other branches of physics. Thus, the celebrated
group $SU(3)$ enters also here when a TQD is subject to an
external magnetic field. And the group $SO(5)$ which plays a role
in the theory of superconductivity is found here when a certain
tuning of the gate voltages in TQD is exercised.

The basic concepts are introduced in section \ref{II}. First, in
sub-section \ref{IIA}, the necessary mathematical ingredients are
introduced, although we try to avoid much rigor. Then, in
sub-section \ref{IIB} we explain how these abstract concepts can
be realized in CQD.  In section \ref{III} the special case of TQD
in the {\it parallel} geometry is discussed at some length. In
subsection \ref{III A} we derive scaling equations for TQD with
even occupation and calculate Kondo temperatures for various
dynamical symmetries. In section \ref{IV} we discuss the physics
of TQD in a {\it series} geometry and point out similarities and
differences between Kondo physics in the two geometries
(sub-section \ref{IV A}). In sub-section \ref{IV B} we concentrate
on the case of even occupation. The dynamical-symmetry phase
diagram is displayed and the experimental consequences are drawn.
The case of odd occupation is exposed in sub-section \ref{IV C}.
Finally, in section \ref{V} a novel phenomena is discussed,
namely, {\it a Kondo effect without a localized spin}. The
anisotropic exchange interaction occurs between the metal electron
spin and the dot Runge-Lenz operator alone. In the conclusions we
underscore the main results obtained here.

\noindent
 The derivation of the pertinent effective spin
Hamiltonians and the establishment of group properties (in
particular identification of the group generators and checking the
corresponding commutation relations) sometimes require lengthy
mathematical expressions. These are collected in the appendices.

\section{Dynamical symmetry of complex quantum dots} \label{II}
\subsection{Generalities} \label{IIA}
In this section we present in some details the concept of
dynamical symmetry, and more particularly, its emergence in CQD.
The term {\it Dynamical Symmetry} implies the symmetry of
eigenvectors of a quantum system forming an irreducible
representations of a certain Lie group. The main ideas and the
relevant mathematical tools can be found, e.g., in Refs.
\onlinecite{Dynsym}. Here they are formulated in a form convenient
for our specific purposes without much mathematical rigor. We have
in mind a quantum system with Hamiltonian $H$ whose eigenstates
$|\Lambda \rangle = |M\mu\rangle$ form ( for a given $M$) a basis
to an irreducible representation of some Lie group ${\cal G}$. The
energies $E_{M}$ do not depend on the "magnetic" quantum number
$\mu$. For definiteness one may think of $M$ as an angular
momentum and of $\mu$ as its projection, so that ${\cal G}$ is
just $SU(2)$. Now let us look for operators which induce
transitions between different eigenstates. An economic way for
identifying them is through the Hubbard operators \cite{Hub}
\begin{equation}
X^{\Lambda\Lambda^\prime}=|\Lambda\rangle\langle \Lambda^\prime|.
\label{Hub}
\end{equation}
It is natural to divide this set of operators into two subsets .
The first one contains the operators $|M\mu\rangle \langle
\mu^{\prime}M| $ while the second one includes operators
$|M\mu\rangle \langle \mu^{\prime}M^ {\prime}| $ for which $|M \mu
\rangle $ and $|M' \mu'\rangle$  belong to a {\it different}
representation space of ${\cal G}$. A central question at this
stage is whether these operators (or rather, certain linear
combinations of them) form a close algebra. In some particular
cases it is possible to form linear combinations within each set
and obtain two new sets of operators $\{S\}$ and $\{R\}$ with the
following properties: 1) For a given $M$ the operators $\{S\}$
generate the $M$ irreducible representation of ${\cal G}$ and
commute with $H$. 2) For a given set $M_{i}$ the operators $\{S\}$
and $\{R\}$ form an algebra (the {\it dynamic} algebra) and
generate a non-compact Lie group ${\cal A}$. The reason for the
adjective {\it dynamic} is that, originally,  the operators
$\{R\}$ do not appear in the bare Hamiltonian $H$ and emerge only
when additional interaction (e.g., dot-lead tunneling) is present.
In the special case ${\cal G}=SU(2)$ the operators in $\{S\}$ are
the vector ${\bf S}$ of spin operators determining the
corresponding irreducible representations,  while the operators in
the set $\{R\}$ can be grouped into a sequence ${\bf R}_{n}$ of
vector operators describing transitions between states belonging
to different representations of the $SU(2)$ group.

 Strictly speaking, the group ${\cal A}$ is not a symmetry group of the
Hamiltonian $H$ since the operators $\{R\}$ do not commute with
$H$. Indeed, let us express $H$ in terms of diagonal Hubbard
operators,
\begin{equation} H=\sum_{\Lambda=M\mu}E_{\Lambda}
|\Lambda\rangle\langle \Lambda| =\sum_{\Lambda}E_{M} X^{\Lambda
\Lambda} ~, \label{HX}
\end{equation}
 so that
\begin{equation}\label{comm}
[X^{\Lambda\Lambda^{\prime}},H]=-(E_{M}-E_{M^{\prime}})X^{\Lambda
\Lambda^{\prime}}.
\end{equation}
As we have mentioned above, the symmetry group ${\cal G}$ of the
Hamiltonian $H$, is generated by the operators
$X^{\Lambda=M\mu,\Lambda^{\prime}=M\mu^{\prime}} $. Remarkably,
however, the dynamics of CQD in contact with metallic leads and/or
an external magnetic field leads to renormalization of the
energies $\{E_{M}\}$ in such a way that a few levels at the bottom
of the spectrum become degenerate, $E_{M_{1}}=E_{M_{2}}=\ldots
E_{M_{n}}$. Hence, in this low energy subspace, the group ${\cal
A}$ generated by the operators $\{S\}$ and $\{R\}$ is a symmetry
group of $H$ referred to as the dynamical symmetry group. The
symbol $R$ is due to the analogy with the Runge-Lenz operator, the
hallmark of dynamical symmetry of the Kepler and Coulomb problems
\cite{foot1}. Below we will use the term dynamical symmetry also
in cases where the levels are not strictly degenerate, their
difference is bounded by a certain energy scale, which is the
Kondo energy in our special case. In that sense, the symmetry is
of course not exact, but rather approximate.

Using the notions of dynamical symmetry, numerous familiar quantum
objects, such as hydrogen atom, quantum oscillator in
$d$-dimensions, quantum rotator, may be described in a compact and
elegant way. We are interested in a special application of this
theory, when the symmetry of the quantum system is approximate and
its violation may be treated as a perturbation. This aspect of
dynamical symmetry was first introduced in particle physics
\cite{DGN}, where the classification of hadron eigenstates is
given in terms of non-compact Lie groups. In our case, the
rotationally invariant object is an isolated quantum dot, whose
spin symmetry is violated by electron tunneling to and from the
leads under the condition of strong Coulomb blockade.
\subsection{Realization in CQD} \label{IIB}
The special cases ${\cal G}=SU(2)$ and ${\cal A}=SO(n)$ or $SU(n)$
is realizable in CQD. Let us first recall the manner in which the
spin vectors appear in the effective low energy Hamiltonian of the
QD in tunneling contact with metallic leads. When strong Coulomb
blockade completely suppresses charge fluctuations in QD, only
spin degrees of freedom are involved in tunneling via the Kondo
mechanism \cite{Glaz}. An {\it isolated} SQD in this regime is
represented solely by its spin vector $\bf{S}$. This is a
manifestation of rotational symmetry which is of geometrical
origin. The exchange interaction $J {\bf s} \cdot {\bf S}$ ({\bf
s} is the spin operator of the metallic electrons) induces
transitions between states belonging to the same spin (and breaks
$SU(2)$ invariance). On the other hand, the low energy spectrum of
spin excitations in CQD is not characterized solely by its spin
operator since there are states close in energy, which belong to
different representation spaces of $SU(2)$. Incidentally, these
might have either the same spin ${\bf S}$ (like, {\it e.g}, in two
different doublets) or a different spin (like, {\it e.g.}, in the
case of singlet-triplet transitions). The exchange interaction
must then contain also other operators ${\bf R}_n$ (the
R-operators mentioned in the previous sections) inducing
transitions between states belonging to different representations.
The interesting physics occurs when the operators ${\bf R}_n$
``approximately'' commute with the Hamiltonian $H_{dot}$ of the
isolated dot. In accordance with our previous discussion, the
R-operators are expressible in terms of Hubbard operators and have
only non-diagonal matrix elements in the basis of the eigenstates
of $H_{dot}$. The spin algebra is then a subalgebra of a more
general non-compact Lie algebra formed by the whole set of vector
operators $\{{\bf S}, {\bf R}_n \}$. This algebra is characterized
by the commutation relations,
\begin{eqnarray}
&&[S_i,S_j]=it_{ijk}S_k,\ \ \ \
[S_i,R_{nj}] = i t_{ijk} R_{nk},\nonumber\\
&&[R_{ni},R_{nj}]  =  i t^n_{ijk} S_k, \label{1.1}
\end{eqnarray}
with structure constants $t_{ijk}$, $t^n_{ijk}$ (here $ijk$ are
Cartesian indices). The R-operators are orthogonal to $\bf S$,
\begin{equation}\label{orth}
{\bf S}\cdot {\bf R}_n=0.
\end{equation}
In the general case, CQDs possess also other symmetry elements
(permutations, reflections, finite rotations). Then, additional
scalar generators $A_p$ arise. These generators also may be
expressed via the bare Hubbard operators, and their commutation
relations with R-operators have the form
\begin{equation}
[R_{ni},R_{mj}] = i g_{ij}^{nmp} A_p,~~
[R_{ni},A_p] = i f_{ij}^{nmp} R_{mj}, \label{1.2}
\end{equation}
with structure constants $g_{ij}^{nmp}$ and $f_{ij}^{nmp}$ ($n\neq
m$). The operators obeying the commutation relations (\ref{1.1})
and (\ref{1.2}) form an $o_n$ algebra. The Casimir operator for
this algebra is
\begin{equation}\label{1.K}
{\cal K}= {\bf S}^2 + \sum_n {\bf
R}_n^2 + \sum_p A_p^2~.
\end{equation}
Various representations of
all these operators via basic Hubbard operators will be
established in the following sections, where the properties of
specific CQDs are studied.

 Next, we show how the dynamical symmetry of CQD is revealed in
the effective spin Hamiltonian describing Kondo tunneling. This
Hamiltonian is derived from the generalized Anderson Hamiltonian
\begin{equation}
H_A=H_{dot}+ H_{lead}+ H_{tun}. \label{1.3}
\end{equation}
The three terms on the right hand side are the dot, lead and
tunneling Hamiltonians, respectively. In the generic case, a
planar CQD is a confined region of a semiconductor, with
complicated multivalley structure secluded between drain and
source leads. The CQD contains several valleys numbered by index
$a$. Some of these valleys are connected with each other by tunnel
channels characterized by coupling constants $W_{aa'}$, and some
of them are connected with the leads by tunneling. The
corresponding tunneling matrix elements are $V_{ab}$ $(b=s,d$
stands for source and drain, respectively). The total number of
electrons $N$ in a {\it neutral} CQD as well as the partial
occupation numbers $N_a$ for the separate wells are regulated by
Coulomb blockade and gate voltages $v_{ga}$ applied to these
wells, with $N=\sum_a N_a$. It is assumed that the capacitive
energy for the whole CQD is strong enough to suppress charged
states with $N'=N\pm1$, which may arise in a process of lead-dot
tunneling.

If the inter-well tunnel matrix elements $W_{aa'}$ are larger than
the dot-lead ones $V_{ab}$ (or if all tunneling strengths are
comparable), it is convenient first to diagonalize $H_{dot}$ and
then consider $H_{tun}$ as a perturbation. In this case $H_{dot}$
may be represented as
\begin{equation}\label{1.4}
H_{dot}=\sum_{\Lambda\in N}E_{\Lambda} |\Lambda\rangle\langle
\Lambda| + \sum_{\lambda\in N\pm 1}E_{\lambda}
|\lambda\rangle\langle \lambda| .
\end{equation}
Here all intradot interactions are taken into account. The kets
$|\Lambda\rangle \equiv |N,q \rangle$ represent eigenstates of
$H_{dot}$ in the charge sector $N$ and quantum numbers $q$,
whereas the kets $|\lambda\rangle \equiv |N\pm 1, p \rangle$ are
eigenstates in the charge sectors $N\pm 1$ with quantum numbers
$p$.  All other charge states are suppressed by Coulomb blockade.
Usually, $q$ and $p$ refer to spin quantum numbers but sometimes
other specifications are required (see below).

The lead Hamiltonian takes a form
\begin{equation}\label{1.5}
H_{lead}= \sum_{k,\alpha,\sigma} \varepsilon_{k
\alpha}c^\dagger_{\alpha k\sigma}c_{\alpha k\sigma}.
\end{equation}
In the general case, the individual dots composing the CQD are
spatially separated, so one should envisage the situation when
each dot is coupled by its own channel to the lead electron
states. So, the electrons in the leads are characterized by the
index $\alpha$, which specifies the lead (source or drain) and the
tunneling channel, as well as by the wave vector $k$ and spin
projection $\sigma$.

The tunnel Hamiltonian involves electron transfer between the
leads and the CQD, and thus couples states $|\Lambda \rangle$ of
the dot with occupation $N$ and states  $|\lambda \rangle$ of the
dot with occupation $N \pm 1$. This is best encoded in terms of
non-diagonal dot Hubbard operators, which  intermix the states
from different {\it charge sectors}
\begin{equation}
 X^{\Lambda
\lambda}=|\Lambda \rangle \langle \lambda|, ~~~ X^{\lambda
\Lambda}=|\lambda \rangle \langle \Lambda|. \label{Xnondiag}
\end{equation}
Thus,
\begin{eqnarray}\label{1.6}
H_t  &=&\sum_{k\alpha \sigma} \sum_{\lambda\in N+1,\Lambda\in
N}\left(V_{\alpha \sigma}^{\Lambda\lambda}c^\dagger_{\alpha
k\sigma}|\Lambda\rangle
\langle \lambda| + H.c. \right ) \nonumber \\
& +&\sum_{k\alpha a\sigma}\sum_{\lambda\in N-1,\Lambda\in N}\left(
V_{\alpha \sigma}^{\lambda\Lambda}c^\dagger_{\alpha
k\sigma}|\lambda\rangle \langle\Lambda| + H.c. \right ),
\end{eqnarray}
where $V_{\alpha  \sigma}^{\lambda\Lambda}=V_{\alpha }\langle
\lambda|d_{a\sigma}|\Lambda\rangle$.

Before turning to calculation of CQD conductance, the relevant
energy scales should be specified. First, we suppose that the
bandwidth of the continuum states in the leads, $D_\alpha$,
substantially exceeds the tunnel coupling constants, $D_\alpha \gg
W_{aa'},V_{\alpha }$ (actually, we consider leads made of the same
material with $D_{as}=D_{ad}=D_0$). Second, each well $a$ in the
CQD is characterized by the "activation energy" defined as
$\Delta_a = E_\Lambda(N_a)-E_\lambda(N_a-1)$, i.e., the energy
necessary to extract one electron from the well containing $N_a$
electrons and move it to the Fermi level of the leads (the Fermi
energy is used as the reference zero energy level from now on).
Note that $\Delta_a$ is tunable by applying the corresponding gate
voltage $v_{ga}$. We are mainly interested in situations where the
condition
\begin{equation}\label{1.8} \Delta_c \sim D_0,~Q_c,
\end{equation}
is satisfied at least for one well labeled by the index $c$. Here
$Q_c$ is a capacitive energy, which is predetermined by the radius
of the well $c$. Eventually, this well with the largest charging
energy is responsible for Kondo-like effects in tunneling,
provided the occupation number $N_c$ is odd. The third condition
assumed in our model is a weak enough Coulomb blockade in all
other wells except that with $a=c$, i.e., $Q_a\ll Q_c$. Finally,
we demand that
\begin{equation}\label{1.7}
b_{\alpha a}\equiv \frac{V_{\alpha }}{\Delta_a}\ll 1,
\end{equation}
for those wells, which are coupled with metallic
leads, and
\begin{equation}\label{1.9}
\beta_{a}=\frac{W_{ac}}{E_{ac}}\ll 1.
\end{equation}
Here $E_{ac}$ are the charge transfer energies for electron
tunneling from the $c$-well to other wells in the CQD.

The interdot coupling under Coulomb blockade in each well
generates indirect exchange interactions between electrons
occupying different wells. Diagonalizing the dot Hamiltonian for a
\textit{given} $N=\sum_a N_a$, one easily finds that the low-lying
spin spectrum in the charge sectors with even occupation $N$
consists of singlet/triplet pairs (spin $S=0$ or 1, respectively).
In charge sectors with odd $N$ the manifold of spin states
consists of doublets and quartets (spin $S$=1/2 and 3/2,
respectively).

The resonance Kondo tunneling is observed as a temperature
dependent zero bias anomaly in tunnel conductance \cite{KKK}.
According to existing theoretical understanding, the quasielastic
cotunneling accompanied by the spin flip transitions in a quantum
dot is responsible for this anomaly. To describe the cotunneling
through a neutral CQD with given $N$, one should integrate out
transitions involving high-energy states from charge sectors with
$N'=N\pm 1$. In the weak coupling regime at $T>T_K$ this procedure
is done by means of perturbation theory which can be employed in a
compact form within the renormalization group (RG) approach
formulated in Refs. \onlinecite{Anderson,Hald}.

As a result of the RG iteration procedure, the energy levels
$E_\Lambda$ in the Hamiltonian (\ref{1.4}) are renormalized and
indirect exchange interactions between the CQD and the leads
arise. The RG procedure is equivalent to summation of the
perturbation series at $T>T_K$, where $T_K$ is the Kondo energy
characterizing the crossover from a perturbative weak coupling
limit to a non-perturbative strong coupling regime. The leading
logarithmic approximation of perturbation theory corresponds to a
single-loop approximation of RG theory. Within this accuracy the
tunnel constants $W$ and $V$ are not renormalized, as well as the
charge transfer energy $\Delta_c$ (\ref{1.8}). Reduction of the
energy scale from the initial value $D_0$ to a lower scale $\sim
T$ results in renormalization of the energy levels $E_\Lambda \to
\bar{E}_\Lambda(D_0/T)$ and generates an indirect exchange
interaction between the dot and the leads with an
(antiferromagnetic) exchange constant ${\it J}$.

The rotational symmetry of a {\it simple} quantum dot is broken by
the spin-dependent interaction with the leads, which arises in
second order in the tunneling amplitude $V_\alpha$. In complete
analogy, the {\it dynamical symmetry} of a {\it composite} quantum
dot is exposed (broken) as encoded in the effective exchange
Hamiltonian. In a generic case, there are, in fact, several
exchange constants arranged within an exchange matrix $\textsf{J}$
which is non-diagonal both in dot and lead quantum numbers. The
corresponding exchange Hamiltonian is responsible for spin-flip
assisted cotunneling through the CQD as well as for
singlet-triplet transitions.

The precise manner in which these statements are quantified will
now be explained. After completing the RG procedure, one arrives
at an effective (or renormalized) Hamiltonian $\bar{H}$ in a
reduced energy scale $\bar{D}$,
\begin{equation}\label{1.12a}
\bar{H} =\bar{H}_{dot}+\bar{H}_{lead}+\bar{H}_{cotun},
\end{equation}
where the effective dot Hamiltonian (\ref{1.4}) is
reduced to
\begin{equation}\label{1.12b}
\bar{H}_{dot}=\sum_{\Lambda\in N} \bar{E}_\Lambda
X^{\Lambda\Lambda}
\end{equation}
written in terms of {\it diagonal} Hubbard operators,
\begin{eqnarray}
&& X^{\Lambda\Lambda}=|\Lambda\rangle\langle
\Lambda|. \label{Xdiag}
\end{eqnarray}
 At this stage, the manifold $\{\Lambda\}\in N$
contains only the renormalized low-energy states within the energy
interval comparable with $T_K$ (to be defined below). Some of
these states may be quasi degenerate, with energy differences
$|\bar{E}_\Lambda- \bar{E}_{\Lambda'}|< T_K$. However, $T_K$
itself is a function of these energy distances (see, e.g.,
\cite{Magn,Pust,Eto02}), and all the levels, which influence
$T_K$, should be retained in (\ref{1.12b}).

The effective cotunneling Hamiltonian acquires the form
\begin{equation}\label{1.11}
H_{cot}= \sum_{\alpha\alpha'}\left(
J_0^{\alpha\alpha'}{\bf S}\cdot {\bf s}^{\alpha\alpha'} + \sum_n
J_n^{\alpha\alpha'}{\bf R}_n\cdot {\bf s}^{\alpha\alpha'}\right).
\end{equation}
Here ${\bf S}$ is the spin operator of CQD in its ground state,
the operators ${\bf s}^{\alpha\alpha'}$ represent the spin states
of lead electrons,
\begin{equation}\label{1.10}
{\bf s}^{\alpha\alpha'}=\frac{1}{2}\sum_{kk'}\sum_{\sigma \sigma'}
c^\dag_{\alpha k\sigma} \hat{\tau}_{\sigma \sigma'} c_{\alpha' k'
\sigma'}~,
\end{equation}
where $\hat{\tau}$ is the vector of Pauli matrices. In the
conventional Kondo effect the logarithmic divergent processes
develop due to spin reversals given by the first term containing
the operator ${\bf S}$. In CQD possessing dynamical symmetry, all
R-vectors are involved in Kondo tunneling. In the following
sections we will show how these additional processes are
manifested in resonance Kondo tunneling through CQD. Note that the
elements of the matrix $\textsf{J}$ are also subject to
temperature dependent renormalization $J_{n}^{\alpha \alpha' } \to
J_{n}^{\alpha \alpha' }(D_0/T)$.

The cotunneling Hamiltonian (\ref{1.11}) is the natural
generalization of the conventional Kondo Hamiltonian $J {\bf s}
\cdot {\bf S}$ for CQDs possessing dynamical symmetries.  In many
cases there are several dot spin 1 operators depending on which
pair of electrons is ``active''. In this pair, one electron sits
in well $c$ and the other one sits in some well $a$. The other
$N-2$ electrons are paired in singlet states. This scenario
applies if $N$ is even. The spin 1 operator for the active pair is
denoted as ${\bf S}_{a}$. (In some sense, the need to specify
which pair couples to $S=1$ while all other pairs are coupled to
$S=0$ is the analog of the seniority scheme in atomic and nuclear
physics (see, e.g., \cite{Talmi})). The cotunneling Hamiltonian
for CQD contains exchange terms  $J_0^{\alpha\alpha'}{\bf
S}_{a}\cdot {\bf s}^{\alpha\alpha'}$. Then, instead of a single
exchange term (first term on the RHS of Eq. (\ref{1.11})), one has
a sum $\sum_a J_a^{\alpha \alpha'}{\bf S}_a\cdot {\bf s}^{\alpha
\alpha'}$. Additional symmetry elements (finite rotations and
reflections) turn the cotunneling Hamiltonian even more
complicated. In the following sections we will consider several
examples of such CQDs. It is seen from (\ref{1.11}), that in the
generic case, both spin and R-vectors may be the sources of
anomalous Kondo resonances. The contribution of these vectors
depends on the hierarchy of the energy states in the manifold. In
principle, it may happen that the main contribution to the Kondo
tunneling is given not by the spin of the dot, but by one of the
R-vectors.

Thus, we arrive at the conclusion that the regular procedure of
reducing the full Hamiltonian of a quantum dot in junctions with
metallic leads to an effective Hamiltonian describing only spin
degrees of freedom of this system reveals a rich dynamical
symmetry of CQD. Strictly speaking, only an isolated QD with $N=1$
is fully described by its spin 1/2 operator obeying $SU(2)$
symmetry without dynamical degrees of freedom. Yet even the doubly
occupied dot with $N=2$ possesses the dynamical symmetry of a {\it
spin rotator} because its spin spectrum consists of a singlet
ground state (S) and a triplet excitation (T). Therefore, an
R-vector describing S/T transitions may be introduced, and the
Kondo tunneling through a dot of this kind may involve spin
excitation under definite physical conditions, e.g., in an
external magnetic field \cite{Magn}. A two-electron quantum dot
under Coulomb blockade constitutes apparently the simplest
non-trivial example of a nano-object with dynamical symmetry of a
spin rotator possessing an $SO(4)$ symmetry.

Dynamical symmetries $SO(n)$ of CQDs are described by non-compact
semi-simple algebras \cite{Cahn}. This non-compactness implies
that the corresponding algebra $o_n$ may be presented as a direct
sum of subalgebras, e.g., $o_4 = o_3 \oplus o_3$. Therefore, the
dynamical symmetry group may be represented as a direct product of
two groups of lower rank. In case of spin rotator the product is
$SO(4)=SU(2)\otimes SU(2)$. Generators of these subgroups may be
constructed from those of the original group. The $SO(4)$ group
possesses a single R-operator ${\bf R}$, and the direct product is
realized by means of the transformation
\begin{equation}\label{1.14}
{\bf K}= \frac{{\bf S}+\bf R}{2},~~~{\bf N}=
\frac{{\bf S}-\bf R}{2}.
\end{equation}
 Both vectors ${\bf K}$ and
${\bf N}$ generate $SU(2)$ symmetry and may be treated as
fictitious S=1/2 spins \cite{Pust}. In some situations these
vectors are real spins localized in different valleys of CQD. In
particular, the transformation (\ref{1.14}) maps a single site
Kondo problem for a DQD possessing $SO(4)$ symmetry to a two-site
Kondo problem for spin 1/2 centers with an $SU(2)$ symmetry (see
discussion in Refs. \onlinecite{KA01,KA02}). For groups of higher
dimensionality ($n\geq 4)$ one can use many different ways of
factorization, which may be represented by means of different
Young tableaux (see Appendix D).

Even in the case $n=4,$ the transformation (\ref{1.14}) is not the
only possible "two-spin" representation. An alternative
representation is realized in an external magnetic field
\cite{KA02}. When the ground state of S/T manifold is a singlet
(the energy $\delta=E_T-E_S>0)$, the Zeeman splitting energy of a
triplet in an external magnetic field may exactly compensate the
exchange splitting $\delta$. This occasional degeneracy is
described by the pseudospin 1/2 formed by the singlet and the up
projection of spin 1 triplet. Two other projections of the triplet
form the second pseudospin 1/2. The Kondo effect induced by
external magnetic field observed in several nano-objects
\cite{ext}, was the first experimental manifestation of dynamical
symmetry in quantum dots.

We outlined in this section the novel features which appear in
effective Kondo Hamiltonians due to the dynamical symmetry of CQD
exhibiting Kondo tunneling. In the following sections we will see
how the additional terms in the Hamiltonian (\ref{1.11}) influence
the properties of Kondo resonance in various structures of CQDs.

\section{Triple quantum dot in parallel geometry} \label{III}

\begin{figure}[htb]
\centering
\includegraphics[width=80mm,height=80mm,angle=0]{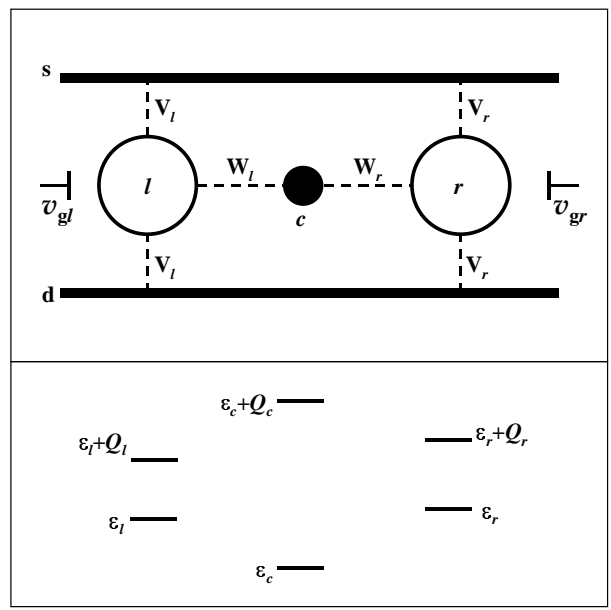}
\caption{Triple quantum dot in parallel geometry and energy levels
of each dot $\varepsilon_{a}=\epsilon_{a}-v_{ga}$ \label{TQD}
(bare energy minus gate voltage).}
\end{figure}

So far we have briefly mentioned a simple structure of CQD, i.e.,
double quantum dot with occupation $N=2$ and employed it to
describe some generic properties of CQD enumerated in the previous
section. This kind of an artificial molecule is the analog of a
hydrogen molecule in the Heitler-London limit \cite{KA01,KA02},
and its $SO(4)$ symmetry reflects the spin properties of
ortho/parahydrogen. A much richer artificial object is a triple
quantum dot (TQD), which can be considered as an analog of a {\it
linear molecule} RH$_2$. The central $(c)$ dot is assumed to have
a smaller radius (and, hence, larger capacitive energy $Q_c$) than
the left $(l)$  and right $(r)$ dots, i.e., $Q_c \gg Q_{l,r}$.
Fig. \ref{TQD} illustrates this configuration in a parallel
geometry, where the "left-right" ($l-r$) reflection plane of the
TQD is perpendicular to the "source-drain" ($s-d$) reflection
plane of metallic electrodes.

To regulate the occupation of TQD as a whole and its constituents
in particular, there is a couple of gates $v_{gl},v_{gr}$ applied
to the $l,r$ dots. The energy levels of single- and two-electron
states in each one of the three constituent dots are shown in the
lower panel of Fig. 1. Here the gate voltages $v_{gl,r}$ are
applied in such a way that the one-electron level $\epsilon_c$ of
a $c$-dot is essentially deeper than those of the $l,r$-dots, so
that the condition (\ref{1.8}) is satisfied for the $c$ dot,
whereas the inequalities (\ref{1.7}) and (\ref{1.9}) are satisfied
for the "active" $l$ and $r$ dots. Tunneling between the side dots
$l,r$ and the central one $c$ with amplitudes  $W_{l,r}$
determines the low energy spin spectrum of the isolated TQD once
its occupation $N$ is given. This system enables the exposure of
much richer
possibilities for additional degeneracy relative
to the DQD setup mentioned
above due to the presence of two channels $(l,r)$.

The full diagonalization procedure of the Hamiltonian $H_{dot}$
for the TQD is presented in Appendix \ref{diag}. When the
condition (\ref{1.9}) is valid, the low-energy manifold for $N=4$
is composed of two singlets $|S_l\rangle,|S_r\rangle$, two
triplets $|T_a\rangle=|\mu_a\rangle$ ($a=l,r,
\mu_a=1_a,0_a,\bar{1}_a$) and a charge transfer singlet exciton
$|Ex\rangle$ with an electron removed from the $c$-well to the
"outer" wells. Within the first order in $\beta_a\ll1$ the
corresponding energies are,
\begin{eqnarray}
E_{S_{a}} &=& {\epsilon}_c +{\epsilon}_{a}+2{\epsilon}_{\bar a}
+Q_{\bar a} -2W_{a}\beta_{a},\nonumber\\
E_{{T_a}} &=& {\epsilon}_c +{\epsilon}_{a}+2{\epsilon}_{\bar
a}+Q_{\bar a} ,
\label{En} \\
E_{Ex} &=& 2{\epsilon}_{l} +2{\epsilon}_{r}+Q_l+Q_r
+2W_{l}\beta_{l}+ 2W_{r}\beta_{r}, \nonumber
\end{eqnarray}
where the charge transfer energies in Eq.(\ref{1.9}) (for
determining $\beta_a$) are $E_{ac}=Q_a+\epsilon_a-\epsilon_c$; the
notation $a=l,r$ and ${\bar a}=r,l$ is used ubiquitously
hereafter.

The completely symmetric configuration,
$\varepsilon_l=\varepsilon_{r}\equiv \varepsilon,~ Q_l=Q_{r}\equiv
Q ,~ W_l=W_{r}\equiv W,$ should be considered separately. In this
case the singlet states form even and odd combinations in close
analogy with the molecular states $\Sigma^\pm$ in axisymmetric
molecules. The odd state $S_-$ and two triplet states are
degenerate:
\begin{eqnarray}
&&E_{S^+}=\varepsilon_c+3\varepsilon+Q -4W\beta ,\nonumber
\\
 &&E_{S^-}=E_{T_{a}}=\varepsilon_c+3\varepsilon+Q, \label{degen}
 \\
&& E_{Ex} = 4{\epsilon} +2Q +4W\beta. \nonumber
\end{eqnarray}
Consideration of these two examples provide us with an opportunity
to investigate the dynamical symmetry of CQD.

\subsection{Derivation and solution of scaling equations}\label{III A}
We commence with the case of TQD with even occupation $N=4$
briefly discussed in Ref. \onlinecite{KKA02}. This configuration
is a direct generalization of an asymmetric spin rotator, i.e.,
the double quantum dot in a side-bound geometry \cite{KA01}.
Compared with the asymmetric DQD, this composite dot possesses one
more symmetry element, i.e., the $l-r$ permutation, which, as will
be seen below, enriches the dynamical properties of CQD.

Following a glance at the energy level scheme (\ref{En}), one is
tempted to conclude outright that for finite $W,$ the ground state
of this TQD configuration is a singlet and consequently there is
no room for the Kondo effect to take place. A more attentive study
of the tunneling problem, however, shows that tunneling between
the TQD and the leads opens the way for a rich Kondo physics
accompanied by numerous dynamical symmetries.

Indeed, inspecting the expressions for the energy levels, one
notices that the singlet states $E_{S_a}$ are modified due to
inter-well tunneling, whereas the triplet states $E_{T_a}$ are
left intact. This difference is due to the admixture of the
singlet states with the charge transfer singlet exciton (see
Appendix \ref{diag}). As was mentioned in the previous section,
the Kondo cotunneling in the perturbative weak coupling regime at
$T,\varepsilon > T_K$ is excellently described within RG formalism
\cite{Anderson,Hald}. According to general prescriptions of this
theory, the renormalizable parameters of the effective low-energy
Hamiltonian in a one-loop approximation are the energy levels
$E_\Lambda$ and the effective indirect exchange vertices
$J_{\Lambda\Lambda'}^{\alpha\alpha'}$.

To apply the RG procedure to the Kondo tunneling through TQD, let
us first specify the terms $H_{lead}$ and $H_{tun}$ in the
Anderson Hamiltonian (\ref{1.3}). The most interesting for us are
situations where the accidental degeneracy of spin states is
realized. So we consider geometries where the device {\it as a
whole} possesses either complete or slightly violated $l-r$ axial
symmetry. Then the quantum number $\alpha$ in $H_{lead}$
(\ref{1.5}) contains the lead index ($s,d$) and the channel index
($l,r$). If the axial $l-r$ symmetry is perfect, the two tunneling
channels are independent. To describe weak violation of this
symmetry we introduce weak interchannel hybridization in the
leads. This hybridization is characterized by a constant
$t_{lr}\ll D_{0},$ which is small first due to the angular
symmetry, and second due to significant spatial separation between
the two channels. The wave vector $k$ is assumed to remain a good
quantum number. Then, having in mind that in our model
$\epsilon_{kas}=\epsilon_{kad}\equiv \epsilon_{ka}$, the
generalized Hamiltonian (\ref{1.5}) acquires the form
\begin{equation}\label{tlead}
H_{lead}=\sum_{ k
\sigma}\sum_{b=s,d}\sum_{a=l,r}\left(\epsilon_{ka} n_{abk\sigma} +
t_{lr} c^{\dag}_{abk\sigma}c_{\bar{a}bk\sigma}\right).
\end{equation}
The tunneling Hamiltonian (\ref{1.6}) is written as
\begin{equation}\label{hyb}
H_{tun}=\sum_{\Lambda\lambda}\sum_{k\sigma}\sum_{ab}(V^{\lambda\Lambda}_{ab\sigma}
c^{\dag}_{abk\sigma} X^{\lambda\Lambda}+H.c.).
\end{equation}
We assume below $V_{as}=V_{ad}\equiv V_a$ (see Fig.\ref{TQD}).

\begin{figure}[h]
\centerline{\epsfig
{figure=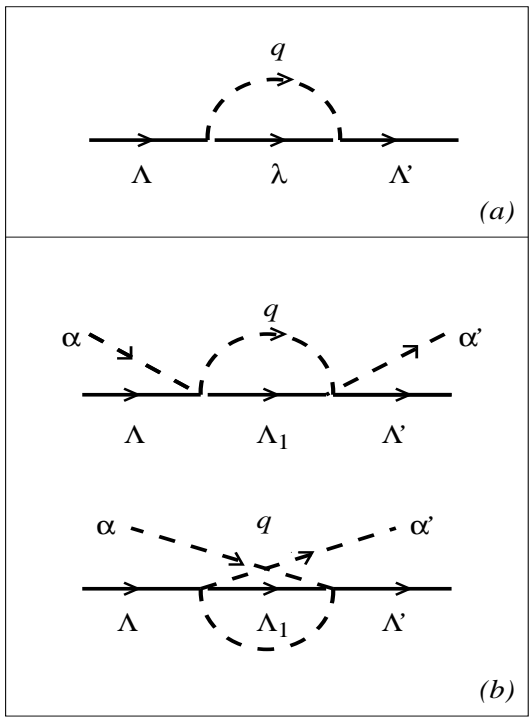,width=70mm,height=95mm,angle=0}} \caption{RG
diagrams for the energy levels $E_\Lambda$ $(a)$ and the effective
exchange vertices $J_{\Lambda\Lambda'}^{\alpha\alpha'}$ $(b)$ (see
text for further explanations).} \label{RGfig}
\end{figure}
The iteration processes, which  characterize the two-step RG
procedure contributing to these parameters are illustrated in
Fig.\ref{RGfig}. The intermediate states in these diagrams are the
high-energy states $|q\rangle$ near the ultraviolet cut-off energy
$D$ of the band continuum in the leads (dashed lines) and the
states $|\lambda\rangle \in N-1$ from adjacent charge sectors,
which are admixed with the low-energy states $|\Lambda\rangle \in
N$ by the tunneling Hamiltonian $H_t$ (\ref{1.6}) (full lines).
For the sake of simplicity we confine ourselves with $N=3$ states
in the charge sector.

In the upper panel, the diagrams contributing to the
renormalization of $H_{dot}$ are shown. In comparison with the
original theory \cite{Hald}, this procedure not only results in
renormalization of the energy levels but also an additional
hybridization of the states $|\Lambda_a\rangle$ via channel mixing
terms in the Hamiltonian (\ref{hyb}). Due to the condition
(\ref{1.8}), the central dot $c$ remains "passive" throughout the
RG procedure.

The mathematical realization of the diagrams displayed in Fig.2a
is encoded in the scaling equations for the energy levels
$E_{\Lambda}$,
\begin{equation}
\pi\,dE_\Lambda/dD=\sum_\lambda
\frac{\Gamma_{\Lambda}}{D-E_{\Lambda\lambda}}. \label{DE}
\end{equation}
Here $E_{\Lambda\lambda}=E_\Lambda-E_\lambda$, $\Gamma_\Lambda$
are the tunnel coupling constants which are different for
different $\Lambda$,
\begin{equation}
\Gamma_{T_a}=\pi \rho_0(V_a^2+2 V_{\bar a}^2),\ \ \ \ \
\Gamma_{S_a}=\alpha_a^2 \Gamma_{T_a}.\label{gam-4}
\end{equation}
Here $\alpha_a=\sqrt{1-2\beta_a^2}$, and $\rho_0$ is the density
of electron states in the leads, which is supposed to be energy
independent. These scaling equations should be solved at some
initial conditions
\begin{equation}\label{init}
E_{\Lambda}(D_{0})=E_{\Lambda}^{(0)},
\end{equation}
where the index $(0)$ marks the bare values of the model
parameters entering the Hamiltonian $H_A$ (\ref{1.3}).

Besides, the diagram of Fig.2a generates a new vertex
$M_{lr}^{\Lambda\Lambda'}$, where the states $\Lambda,\Lambda'$
are either two singlets $S_l,S_{r}$ or two triplets $T_l,T_{r}$.
The Haldane's iteration procedure results in a scaling equation
\begin{equation}
d M_{lr}/d\ln D = - \gamma \label{sclr}
\end{equation}
with the initial condition $M_{lr}(D_0)=0$ and the flow rate
$\gamma=\rho_0^2 V_l V_r t_{lr}.$ After performing the Haldane's
procedure we formally come to the scaled dot Hamiltonian
\begin{equation}\label{scah}
H=\sum_{\Lambda_a}{E}_{\Lambda_a}
X^{\Lambda_a\Lambda_a}+\sum_{\Lambda_{a} \Lambda_{\bar a}}M_{lr}
X^{\Lambda_a \Lambda_{\bar a}}
\end{equation}
with the parameters ${E}_{\Lambda_a}$ and $M_{lr}$ depending on
the running variable $D$.

 Due to the above mentioned dependence of tunneling rates on
the index $\Lambda$, namely the possibility of $\Gamma_T>\Gamma_S$
and $\Gamma_{S_-}>\Gamma_{S_+}$, the scaling trajectories
$E_\Lambda (D)$ may cross at some value of the monotonically
decreasing energy parameter $D$. The nature of level crossing is
predetermined by the initial conditions (\ref{init}) and the
ratios between the tunneling rates $\Gamma_\Lambda$. As long as
the inequality $|E_{\Lambda\lambda}|\ll D$ is effective and all
levels are non-degenerate, the scaling equations (\ref{DE}) may be
approximated as
\begin{equation}
\pi\,dE_\Lambda/d\ln D=\Gamma_\Lambda. \label{DE2}
\end{equation}

The scaling trajectories are determined by the scaling invariants
for equations (\ref{DE}),
\begin{equation}
E_\Lambda^{\ast }=E_{\Lambda}(D)-\pi^{-1} \Gamma _\Lambda\ln(\pi
D/\Gamma _\Lambda), \label{inv}
\end{equation}
tuned to satisfy the initial conditions. With decreasing energy
scale $D$ these trajectories flatten and become $D$-independent in
the so called Schrieffer-Wolff (SW) limit, which is reached when
the activation energies $\Delta_a$ become comparable with $D$. The
corresponding effective bandwidth is denoted as $\bar{D}$ (we
suppose, for the sake of simplicity, that $\Delta_a< Q_a$, so that
only the states $|\lambda\rangle$ with $N'=N-1$ are relevant). The
simultaneous evolution of interchannel hybridization parameter is
described by the solution of scaling equation (\ref{sclr}),
\begin{equation}\label{log}
M_{lr}(\bar{D})=\gamma \ln\frac{D_0}{\bar D}.
\end{equation}

If this remarkable level crossing occurs at $D>\bar{D}$, we arrive
at the situation where {\it adding an indirect exchange
interaction between the TQD and the leads changes the magnetic
state of the TQD from singlet to triplet}. Those states
$E_\Lambda$, which remain close enough to the new ground state are
involved in the Kondo tunneling. As a result, the TQD acquires a
rich dynamical symmetry structure instead of the trivial symmetry
of spin singlet predetermined by the initial energy level scheme
(\ref{En}). Appearance of the logarithmic enhancement of the
hybridization parameter $M_{lr}$ (\ref{log}) does not radically
influence the general picture, provided the flow trajectories
cross far from the SW line , due to a very small hybridization
$\gamma\ll \Gamma_\Lambda\ll D$. However, we are interested just
in cases when the accidental degeneracy occurs at the SW line.
Various possibilities of this degeneracy are considered below.

\begin{figure}[h] \centerline{\epsfig
{figure=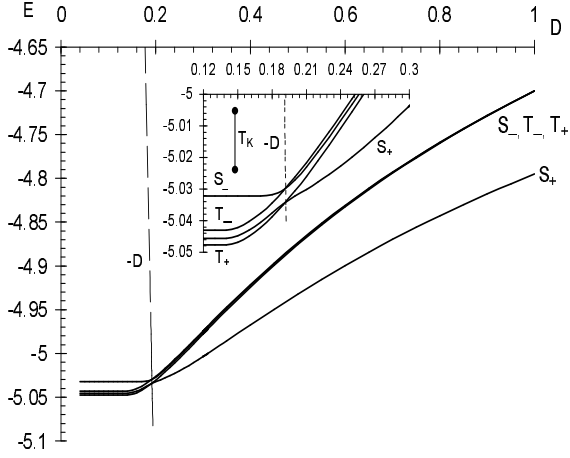,width=75mm,height=60mm,angle=0}} \caption{Scaling
trajectories for $P\times SO(4)\times SO(4)$ symmetry in the SW
regime. Inset: Zoomed in avoided level crossing pattern near the SW line}
\label{pdotso4}
\end{figure}

The flow diagrams leading to a non-trivial dynamical symmetry of
TQD with $N=4$ are presented in
Figs.\ref{pdotso4},\ref{SO5},\ref{SO7}. The horizontal axis on
these diagrams is the current energy scale $D$ for lead electrons,
where the vertical axes represent the energy levels
$E_\Lambda(D)$. The dashed line $E=-D$ establishes the SW boundary
for these levels.

Before turning to highly degenerate situations, where the system
possesses specific $SO(n)$ symmetry, it is instructive to consider
the general case, where all flow trajectories $E_\Lambda(\bar D)$
are involved in Kondo tunneling in the SW limit. This happens when
the whole octet of spin singlets and triplets forming the manifold
(\ref{En}) remains within the energy interval $\sim T_K$ in the SW
limit. The level repulsion effect does not prevent formation of
such multiplet, provided $t_{lr}$ is small enough and the
inequality
\begin{equation}\label{ineq1}
M_{lr}(\bar D)<T_K
\end{equation}
is valid. At this stage, the SW procedure for constructing the
effective spin Hamiltonian in the subspace $\mathbb{R}_8=\{T_l,
S_l, T_r, S_r\}$ should be applied. This procedure excludes the
charged states generated by $H_t$ to second order in perturbation
theory (see, e.g.,\cite{Hewson}).

The effective cotunneling Hamiltonian can be derived using
Schrieffer-Wolf procedure \cite{SW} (see Appendix \ref{H-spin}).
To simplify the SW transformation, one should first rationalize
the tunneling matrix ${\sf V}$ in the Hamiltonian (\ref{hyb}).
This $4\times 4$ matrix is diagonalized in the $s-d, l-r$ space by
means of the transformation to even/odd combinations of lead
electron $k$-states and similar symmetric/antisymmetric
combinations of $l,r$ electrons in the dots. The form of this
transformation for symmetric TQD can be found in Appendix
\ref{GR-trans}. Like in the case of conventional QD \cite{Glaz},
this transformation eliminates the odd combination of $s-d$
electron wave functions from tunneling Hamiltonian.

It should be emphasized that this transformation does  not exclude
the odd component from $H_{tun}$ in case of TQD in a series
geometry \cite{EPL}. The same is valid for the Hamiltonians
(\ref{tlead}), (\ref{hyb}) with $t_{lr}=0$: in this case the
rotation in $s-d$ space conformally maps the Hamiltonian $H_A$
(\ref{1.3}) for TQD in parallel geometry onto that for TQD in
series. Both these cases will be considered in Section \ref{IV}.

 Unlike the case of DQD studied in Refs.
\onlinecite{KA01,KA02}, where the spin operators are the total
spin ${\bf S}$ and a single R-operator, describing S/T
transitions, the TQD is represented by several spin operators
corresponding to different Young tableaux (see Appendix D). To
order $O(|V|^2)$, then,
\begin{eqnarray}
H_{cot}&=&\sum_{\Lambda_a}\bar
{E}_{\Lambda_a}X^{{\Lambda_a}{\Lambda_a}} +\sum_{\Lambda_a
\Lambda_{\bar a}}\bar
{M}_{{lr}}X^{{\Lambda_a}{\Lambda_{\bar a}}}\nonumber\\
 &+&\sum_{ k
\sigma}\sum_{b=s,d}\sum_{a=l,r}\left(\epsilon_{ka} n_{abk\sigma} +
t_{lr} c^{+}_{abk\sigma}c_{\bar{a}bk\sigma}\right)\nonumber\\
&+&
 \sum_{a=l,r}J^T_{a} {\bf S}_{a}\cdot {\bf s}_a+ J_{lr}{\hat P}
\sum_{a=l,r}{\bf S}_{a}\cdot {\bf s}_{{\bar a}a}\label{generic}\\
&+&\sum_{a=l,r}J^{ST}_a {\bf R}_{a}\cdot {\bf s}_a
+J_{lr}\sum_{a=l,r}\tilde{\bf R}_{a}\cdot {\bf s}_{a{\bar
a}}\nonumber.
\end{eqnarray}
Here we recall that $\bar {E}_{\Lambda_a}=E_{\Lambda_a}(\bar{D})$,
$\bar {M}_{lr}=M_{lr}(\bar{D})$, and the effective exchange
constants are
\begin{eqnarray}
J^T_{a}&=&\frac{V_a^2}{\epsilon_F-\epsilon_a},\ \ \ \ \ \
J^{ST}_a=\alpha_aJ^T_a,\nonumber\\
J_{lr}&=&\frac{V_lV_r}{2}\left(\frac{1}{\epsilon_F-\epsilon_l}+\frac{1}
{\epsilon_F-\epsilon_r}\right).\label{J}
\end{eqnarray}
The vector operators ${\bf S}_a,{\bf R}_a,{\bf \tilde{R}}_a$ and
the permutation operator $\hat{P}$ manifest the dynamical symmetry
of TQD in a subspace $\mathbb{R}_8$. The permutation operator
\begin{equation}\label{perm}
{\hat P}=\sum_{a=l,r}\Big(X^{S_a S_{\bar a}}+\sum_{\mu=1,0,{\bar
1} } X^{\mu_a \mu_{\bar a}}\Big)
\end{equation}
commutes with ${\bf S}_l +{\bf S}_r$ and ${\bf R}_l +{\bf R}_r$.

The spherical components of these vectors are defined via Hubbard
operators connecting different states of the octet,
\begin{eqnarray}
S^{+}_a &=& \sqrt{2}( X^{1_a0_a}+X^{0_a\bar{1}_a}),\ \ \ \ \ \ \ \
\ \ S^{-}_a =
(S^{+}_a)^\dagger, \nonumber\\
S^z_{a}&=&X^{1_a1_a}-X^{\bar{1}_a\bar{1}_a}\nonumber\\
R^{+}_a &=& \sqrt{2}( X^{1_aS_a}-X^{S_a\bar{1}_a}),\ \ \ \ \ \ \ \
\   R^{-}_a =
(R^{+}_a)^\dagger,\nonumber\\
R^z_{a} &=&-(X^{0_aS_a}+X^{S_a0_a}),  \nonumber \\
\tilde{R}^{+}_{a} &=& \sqrt{2}(\alpha_{\bar a} X^{1_aS_{\bar
a}}-\alpha_{a} X^{S_a\bar{1}_{\bar a}}),\ \ \tilde{R}^{-}_{a} =
(\tilde{R}^{+}_{a})^\dagger,\nonumber\\
\tilde{R}^z_{a} &=& -(\alpha_{\bar a} X^{0_aS_{\bar a}}+\alpha_{a}
X^{S_a0_{\bar a}}). \label{comm1}
\end{eqnarray}
In addition to the spin operator (\ref{1.10}) for conduction
electrons, new spin operators are required,
\begin{equation}\label{1.100}
{\bf s}_{a \bar a}= \frac{1}{2}\sum_{kk'}\sum_{\sigma \sigma'}
c^\dag_{ak\sigma} \hat{\tau}_{\sigma \sigma'}c_{{\bar a}k'
\sigma'}.
\end{equation}
An extra symmetry element (l-r permutation) results in more
complicated algebra which involves new R-operator ${\bf
\tilde{R}}$ and the permutation operator $\hat{P}$ interchanging
$l$ and $r$ components of TQD.

One can derive from the generic Hamiltonian (\ref{generic}) more
symmetric effective Hamiltonians describing partly degenerate
configurations illustrated by the flow diagrams of Figs.
\ref{pdotso4},\ref{SO5},\ref{SO7}. These are the cases when the
level crossing occurs in a nearest vicinity of the SW line in the
flow diagram. It is important to distinguish between the cases of
generic and accidental symmetry. In the former case the device
possesses intrinsic $l-r$ and $s-d$ symmetry, i.e., the left and
right dots are identical, the corresponding tunnel parameters are
equal, and left and right leads also mirror each other, namely,
$\epsilon_{kl}=\epsilon_{kr}\equiv\epsilon_{k}$. In the latter
case the gate voltages violates $l-r$ symmetry, e.g., make
$\epsilon_{l}\neq \epsilon_{r}$, $V_{l}\neq V_{r}$, etc. The level
degeneracy is achieved due to competition between the $l-r$
interdot tunneling and the lead-dot tunneling without changing the
symmetry of the Hamiltonian.

The basic spin Hamiltonian (\ref{generic}) acquires a more compact
form, when a TQD possesses generic or accidental degeneracy. In
these cases the operators (\ref{comm1}) form close algebras, which
predetermine the dynamical symmetry of Kondo tunneling.  We start
the discussion of the pertinent $SO(n)$ symmetries with the most
degenerate configuration (Fig. \ref{pdotso4}), where the TQD
possesses generic $l-r$ axial symmetry, i.e., the left and right
dots are completely equivalent. Then the energy spectrum of an
isolated TQD is given by Eqs.(\ref{degen}). The four-electron wave
functions are calculated in Appendix (\ref{diag}). Such TQD is a
straightforward generalization of the so called T-shaped DQD
introduced in Refs. \onlinecite{KA01,KA02,Tshap}. It is clear,
that attachment of a third dot simply adds one more element to the
symmetry group $SO(4)$, namely the $l-r$ permutation $\hat P$,
which is parity sensitive.

To reduce the Hamiltonian (\ref{generic}) into a more symmetric
form, we rewrite the Hubbard operators in terms of new eigenstates
$\bar E_\Lambda$, recalculated with account of generic degeneracy
(\ref{degen}) and $l-r$ mixing $\bar M_{lr}$. In assuming that the
latter coupling parameter is the smallest one, it results in
insignificant additional remormalization  $\sim \mp|\bar
M_{lr}|^2/(\varepsilon +Q-\varepsilon_c)$ of the states $E_{S+}$
and $E_{Ex}$. Besides, it intermixes the triplet states and
changes their nomenclature from left/right to even/odd. The
corresponding energy levels are
\begin{equation}\label{lre}
E_{T\pm}(\bar D)=E_{T_a}\mp \bar M_{lr}.
\end{equation}

The flow trajectories for two pairs of states $(T_+, T_-)$ and
$(S_+, S_-)$ diverge slowly with decreasing $D$. If this
divergence is negligible in the scale of $T_K$, then three nearly
coincident trajectories $E_{T\pm},E_{S_-}$ cross the fourth
trajectory $E_{S+}$ at some point, since the inequality
$\Gamma_{S_+}<\Gamma_{T\pm}=\Gamma_{S-}$ with $\Gamma_{T\pm}=3\pi
\rho_o V^2,\ \ \Gamma_{S_+}=\alpha\Gamma_{T\pm}$ is valid
($\alpha=\sqrt{1-4\beta^2}$). If this level crossing happens near
the SW line, we arrive at a case of complete degeneracy of the
renormalized spectrum, and the whole octet $\mathbb{R}_8$ is
involved in the dynamical symmetry (Fig. \ref{pdotso4}). The fine
structure of the flow diagram in the region of avoided level
crossing is shown in the inset.

Since the tunneling occurs in even and odd channels independently,
the parity is conserved also in indirect SW exchange. As a result,
the effective spin Hamiltonian (\ref{generic}) acquires the form
\begin{eqnarray}
&&{H}_{cot} =\sum_{\Lambda_{\eta}}{\bar
E}_{\Lambda_\eta}X^{\Lambda_\eta\Lambda_\eta}
+\sum_{k\sigma}\sum_{\eta=g,u}\epsilon_{k\eta} c^{\dagger}_{\eta
k\sigma }c_{\eta k
\sigma}\nonumber\\
&&+ \sum_{\eta=g,u}J^T_{1\eta} {\bf S}_{\eta}\cdot {\bf s}_\eta+
\sum_{\eta=g,u}J^{ST}_{1\eta} {\bf R}_{\eta}\cdot {\bf s}_\eta
\label{sym-h1-mlr}\\
 &&+J^T_{2}
\sum_{\eta}{\bf S}_{\eta{\bar \eta}}\cdot {\bf s}_{\eta{\bar
\eta}} +\sum_{\eta}(J^{ST}_{2\eta}{\bf R}^{(1)}_{\eta{\bar
\eta}}+J^{ST}_{2{\bar\eta}}{\bf R}^{(2)}_{\eta{\bar \eta}})\cdot
{\bf s}_{\eta{\bar \eta}}.\nonumber
\end{eqnarray}
Here $\epsilon_{kg}=\epsilon_{k}-t_{lr}$,
$\epsilon_{ku}=\epsilon_{k}+t_{lr}$ and the lead operators
$c_{\eta k\sigma}$ $(\eta=g,u)$ are defined in Appendix
\ref{GR-trans}. The operators ${\bf S}_\eta$, ${\bf R}_\eta$ are
defined analogously to ${\bf S}_a$, ${\bf R}_a$ in Eq.(\ref{comm1}),
and the vector operators ${\bf S}_{\eta{\bar \eta}},$ ${\bf
R}^{(1)}_{\eta{\bar \eta}}$, ${\bf R}^{(2)}_{\eta{\bar \eta}}$ are
defined as:
\begin{eqnarray}
{\bf S}_{\eta{\bar \eta}}&=&X^{\eta{\bar \eta}}{\bf S}_{{\bar
\eta}},\ \ \ \ \ {\bf R}^{(1)}_{\eta{\bar \eta}}+{\bf
R}^{(2)}_{\eta{\bar \eta}}=X^{\eta{\bar \eta}}{\bf R}_{{\bar
\eta}}.\label{eta2}
\end{eqnarray}
The spherical components of the operators ${\bf
R}^{(1)}_{\eta{\bar \eta}}$ and ${\bf R}^{(2)}_{\eta{\bar \eta}}$
are given by
\begin{eqnarray}
R^{(1)+}_{\eta{\bar \eta}}&=&-\sqrt{2}X^{S_{\eta}{\bar 1}_{\bar
\eta}}, \ \ \ \ {R}^{(1)-}_{\eta{\bar \eta}}=(R^{(1)+}_{\eta{\bar
\eta}})^{\dag},\nonumber\\
R^{(2)+}_{\eta{\bar \eta}}&=&\sqrt{2}X^{1_{\eta}{S}_{\bar \eta}},
\ \ \ \ \ \; {R}^{(2)-}_{\eta{\bar \eta}}=(R^{(2)+}_{\eta{\bar
\eta}})^{\dag},\nonumber\\
R^{(1)z}_{\eta{\bar\eta}}&=&-X^{S_{\eta}0_{\bar \eta}}, \ \ \ \ \
\ \ \ R^{(2)z}_{\eta{\bar \eta}}=-X^{0_{\eta}S_{\bar
\eta}}.\label{R-eta12}
\end{eqnarray}

The spin operators for the electrons in the leads are introduced
by the obvious relations
\begin{eqnarray}
{\bf s}_{g}&=&\frac{1}{2}\sum_{kk'}\sum_{\sigma \sigma'}
c^\dag_{gk\sigma} \hat{\tau}_{\sigma \sigma'} c_{gk'\sigma'},\nonumber\\
{\bf s}_{u}&=&\frac{1}{2}\sum_{kk'}\sum_{\sigma \sigma'}
c^\dag_{uk\sigma} \hat{\tau}_{\sigma \sigma'} c_{uk'\sigma'},\label{gu-operators}\\
{\bf s}_{gu}&=&\frac{1}{2}\sum_{kk'}\sum_{\sigma \sigma'}
c^\dag_{gk\sigma} \hat{\tau}_{\sigma \sigma'} c_{uk'\sigma'},\ \
{\bf s}_{ug}=({\bf s}_{gu})^{\dag},\nonumber
\end{eqnarray}
instead of (\ref{1.10}). Now the operator algebra is given by the
closed system of commutation relations which is a generalization
of the $o_4$ algebra,
\begin{eqnarray}\label{algebra}
&&\lbrack S_{\eta j},S_{\eta' k}]=ie_{jkm}\delta_{\eta\eta'}S_{\eta m},\nonumber \\
&&[R_{\eta j},R_{\eta' k}]=ie_{jkm}\delta_{\eta\eta'}¥S_{\eta m},\nonumber\\
&&\lbrack R_{\eta j},S_{\eta'
k}]=ie_{jkm}\delta_{\eta\eta'}R_{\eta m}.
\end{eqnarray}
The operators ${\bf S}_\eta$ are orthogonal to ${\bf R}_\eta$, and
the Casimir operators in this case are ${\cal K}_\eta={\bf
S}_\eta^{2}+{\bf R}_\eta^{2}=3.$ This justifies the qualification
of such TQD as a {\it double spin rotator} which is obtained from
the spin rotator considered in Refs. \cite{KA01,KA02} by a mirror
reflection. The symmetry of such TQD is $P\times SO(4)\times
SO(4).$

Four additional vertices appear in the effective spin Hamiltonian
(\ref{sym-h1-mlr}) at the second stage of Haldane-Anderson scaling
procedure \cite{Anderson}. As a result, the exchange part of the
Hamiltonian (\ref{sym-h1-mlr}) takes the form
\begin{eqnarray}
&&{H}_{cot} =\sum_{\eta=g,u}J^T_{1\eta} {\bf S}_{\eta}\cdot {\bf
s}_\eta+ \sum_{\eta=g,u}J^{ST}_{1\eta} {\bf R}_{\eta}\cdot {\bf
s}_\eta\nonumber\\
 &&+J^T_{2}
\sum_{\eta}{\bf S}_{\eta{\bar \eta}}\cdot {\bf s}_{\eta{\bar
\eta}} +\sum_{\eta}(J^{ST}_{2\eta}{\bf R}^{(1)}_{\eta{\bar
\eta}}+J^{ST}_{2{\bar\eta}}{\bf R}^{(2)}_{\eta{\bar \eta}})\cdot
{\bf s}_{\eta{\bar \eta}}\nonumber\\
&&+ \sum_{\eta}J^T_{3\eta} {\bf S}_{\eta}\cdot {\bf s}_{\bar\eta}+
\sum_{\eta}J^{ST}_{3\eta} {\bf R}_{\eta}\cdot {\bf
s}_{\bar\eta}.\label{sym-h-mlr}
\end{eqnarray}

The coupling constants in the Hamiltonian (\ref{sym-h-mlr}) are
subject to renormalization. Their values at $D=\bar D$ are taken
as a boundary conditions
\begin{eqnarray}
J^T_{1\eta}({\bar D})&=&J^T_2({\bar D})=J^{ST}_{1u}({\bar
D})=J^{ST}_{2u}({\bar
D})=\frac{V^2}{\epsilon_{F}-\epsilon},\label{bcond}\\
J^T_{3\eta}({\bar D})&=&J^{ST}_{3u}({\bar D})=0, \
J^{ST}_{ig}({\bar D})=\alpha J^T_{ig}({\bar D}) \
(i=1,2,3)\nonumber
\end{eqnarray}
for solving the scaling equations. These can be written in the
following form:
\begin{eqnarray}
\frac{dj_{1\eta}^{T}}{d\ln d} &=& -\Big[(j^{T}_{1\eta})^2
+2(-1)^\eta m_{lr}j^{T}_{1\eta} \nonumber\\
&+&(j^{ST}_{1\eta})^2
+\frac{(j^{T}_2)^2}{2}+\frac{(j^{ST}_{2{\bar \eta}})^2}{2}\Big],\nonumber\\
\frac{dj_{2}^{T}}{d\ln d} &=&
-\frac{1}{2}\Big[\sum_{\eta=g,u}\{j^{T}_2(j^{T}_{1\eta}+
j^{T}_{3\eta}) + j^{ST}_{2\eta}(
j^{ST}_{1\eta}+j^{ST}_{3\eta})\}\Big],\nonumber\\
\frac{dj_{3\eta}^{T}}{d\ln d} &=& -\Big[(j^{T}_{3\eta})^2
+2(-1)^\eta m_{lr}j^{T}_{3\eta}\nonumber\\
&+&(j^{ST}_{3\eta})^2+
\frac{(j^{T}_2)^2}{2}+\frac{(j^{ST}_{2{\bar \eta}})^2}{2}\Big],\nonumber\\
\frac{dj_{1\eta}^{ST}}{d\ln d} &=&-\Big[2j^{T}_{1\eta}
j^{ST}_{1\eta}+2(-1)^\eta m_{lr}j_{1\eta}^{ST}+ j^{T}_2
j^{ST}_{2\eta} \Big],\nonumber\\
\frac{dj_{2\eta}^{ST}}{d\ln d} &=&
-\frac{1}{2}\Big[\sum_{\eta}j^{T}_2(j^{ST}_{1\eta}+
j^{ST}_{3\eta})+ 2j^{ST}_{2\eta}
(j^{T}_{1{\bar \eta}}+j^{T}_{3{\bar \eta}}) \Big],\nonumber\\
\frac{dj_{3\eta}^{ST}}{d\ln d} &=&-\Big[2j^{T}_{3\eta}
j^{ST}_{3\eta}+2(-1)^\eta m_{lr}j_{3\eta}^{ST}+ j^{T}_2
j^{ST}_{2\eta} \Big],\label{sc-eq-so4}
\end{eqnarray}
where $j_{i\eta}=\rho_0 J_{i\eta}$ ($i=1,2,3$), $d=\rho_0 D$ and
$m_{lr}=\rho_0 M_{lr}.$

Solution of Eqs. (\ref{sc-eq-so4}) yields the Kondo temperature
\begin{equation}
T_{K0}=\bar{D}\left(1-\frac{8m_{lr}}{(\sqrt{3}+1)(3j_{1g}^{T}+j_{1g}^{ST})}\right)
^{\displaystyle{\frac{1}{2m_{lr}}}}.\label{ Tsym}
\end{equation}
The limiting value of this relation for perfect $l-r$ symmetry of
the leads is
\begin{equation}\label{tko}
\lim_{m_{lr}\to 0}T_{K0} ={\bar D} \exp \left(
-\frac{4}{(\sqrt{3}+1)(3j_{1g}^T+j_{1g}^{ST})} \right).
\end{equation}
Here and below the coupling constants $j_i(D)$ in all equations
for $T_K$ are taken at $D=\bar D$. We see that avoided crossing
effect in the case of slightly violated $l-r$ symmetry of TQD
turns the Kondo temperature to be a function of the level
splitting (\ref{lre}). Similar situation has been noticed in
previous studies of DQD \cite{KA01,KA02} and planar QD with even
occupation \cite{Magn}, where $T_K$ turned out to be a
monotonically decreasing function of S/T splitting energy $
\delta={\bar E}_{S}-{\bar E}_{T}$ with a maximum at $\delta=0$.
Now the Kondo temperature is a function of two parameters,
$T_K(M_{lr},\delta).$
\begin{figure}[h]
\centerline{\epsfig{figure=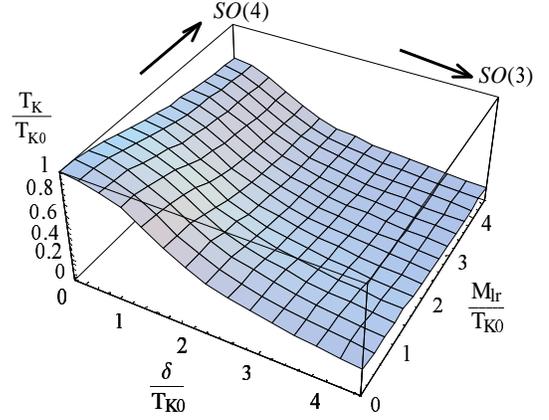,width=70mm,height=55mm,angle=0}}
\caption{Variation of $T_K$ with parameters $\delta$ and $M_{lr}$
(see text for further details).} \label{twopar}
\end{figure}
Looking at Fig.\ref{pdotso4} (which corresponds to $\delta=0$) we
notice that for large enough $M_{lr}$, when the inequality
(\ref{ineq1}) is violated, $M_{lr}\gg T_K$, the symmetry of TQD is
reduced to $SO(4)$ symmetry of S/T manifold with the Kondo
temperature
\begin{equation}\label{TK1}
T_{K1}=\bar{D}\exp\left\{-\frac{1}{j_{1}^T+j_{1}^{ST}}\right\}.
\end{equation}
Additional S/T splitting induced by the gate voltage
($\epsilon_l\neq \epsilon_r$) results in further decrease of $T_K$
as a function of $\delta$. The asymptotic form of the function
$T_K(\delta)$ is
\begin{equation} \label{Tdelta}
\frac{T_K}{T_{K1}}\approx
\left(\frac{T_{K1}}{\delta}\right)^{\alpha}
\end{equation}
(cf. \cite{Magn,Pust}). In the limit of $\delta\rightarrow
\bar{D}$ the singlet state should be excluded from the manifold,
and the symmetry of the TQD with spin one in this case is $SO(3).$
The general shape of $T_K(M_{lr},\delta)$ surface is presented in
Fig.\ref{twopar}. Thus the Kondo effect for the TQD with mirror
symmetry is characterized by the stable infinite fixed point
characteristic for the {\it underscreened} spin one dot, similar
to that for DQD \cite{KA01,KA02}.

Now we turn to asymmetric configurations where $E_{lc}\neq
E_{rc}$, ${\Gamma_{T_r}}\neq{\Gamma_{T_l}}$. In this case the
system loses the $l-r$ symmetry, and it is more convenient to
return to the initial variables used in the generic Hamiltonian
(\ref{generic}).
\begin{figure}[h]
\centerline{\epsfig{figure=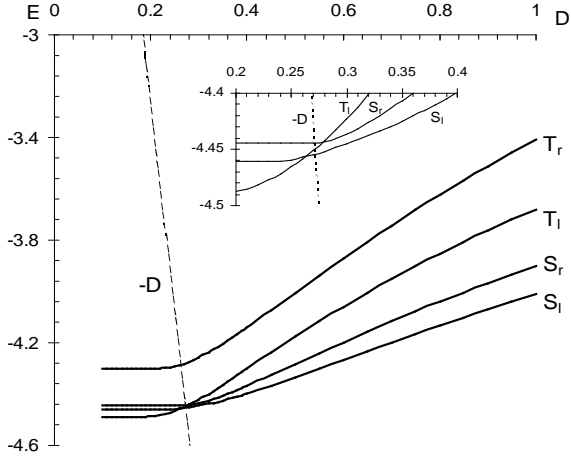,width=75mm,height=60mm,angle=0}}
\caption{Scaling trajectories resulting in an $SO(5)$ symmetry in
the SW regime. Inset: Zoomed in avoided level crossing pattern near the
SW line.} \label{SO5}
\end{figure}

When the Haldane renormalization results in {\it accidental}
degeneracy of two singlets and one triplet, ${\bar E}_{S_l}\approx
{\bar E}_{T_l}\approx {\bar E}_{S_r}< {\bar E}_{T_r}$ (Fig.
\ref{SO5}), the TQD acquires an $SO(5)$ symmetry of a manifold
$\{T_l,S_l,S_r\}$. In this case the SW Hamiltonian (\ref{generic})
transforms into
\begin{eqnarray}
H&=&\sum_{\Lambda=T_l,S_l,S_r}{\bar{E}}_{\Lambda}X^{\Lambda\Lambda}
+M_{lr}(X^{S_lS_r}+X^{S_rS_l})\label{sym5-tlr}\\
&+&\sum_{ k \sigma}\sum_{a=l,r}\epsilon_{ka}
c^{+}_{ak\sigma}c_{ak\sigma}+t_{lr}\sum_{ k \sigma}\sum_{a=l,r}
c^{+}_{ak\sigma }c_{{\bar a}k\sigma }
\nonumber\\
&+&J_{1}{\bf S}_{l}\cdot {\bf s}_l+J_{2}{\bf R}_{l}\cdot {\bf
s}_l
+ J_3(\tilde{\bf R}_{1}\cdot {\bf s}_{rl}+\tilde{\bf R}_{2}\cdot
{\bf s}_{lr}),\nonumber
\end{eqnarray}
where $J_1=J^T_{l}$, $J_2=J^{ST}_{l}$ and $J_3=\alpha_r J_{lr}$.
The spherical components of the vector operators $\tilde{\bf R}_1$
and $\tilde{\bf R}_2$ are given by the following expressions,
\begin{eqnarray} &&\tilde{R}^{+}_1
=-\sqrt{2}X^{S_r\bar{1}_l},\ \tilde{R}^{-}_1=\sqrt{2}
X^{S_r1_l}, \ \tilde{R}_{1z} =-X^{S_r0_l},\nonumber\\
&&\tilde{R}^+_2=(\tilde{R}^{-}_1)^\dag,\ \ \ \ \ \ \;
\tilde{R}^-_2=(\tilde{R}^{+}_1)^\dag,\ \ \ \ \
\tilde{R}_{2z}=\tilde{R}_{1z}^\dag. \label{r-tild}
\end{eqnarray}
 The group generators of the $o_5$
algebra are the $l$-vectors ${\bf S}_l, {\bf R}_l$  from
(\ref{comm1}) and the operators intermixing $l$- and $r$-states,
namely the vector ${\bf \tilde{R}}={\bf \tilde{R}}_1+{\bf
\tilde{R}}_2$,
\begin{eqnarray}\label{tilt}
&&{\tilde R}^{+}=\sqrt{2}(X^{1_{l}S_{r}}-X^{S_{r}\bar{1}_{l}}),~~
{\tilde R}^{-}=({\tilde R}^{+})^{\dagger},\nonumber\\
&&{\tilde R}^{z}=-(X^{0_{l}S_{r}}+X^{S_{r}0_{l}}),
\end{eqnarray}
and a scalar $A$ interchanging $l,r$ variables of degenerate
singlets
\begin{equation}\label{scal}
A=i(X^{S_rS_l}-X^{S_lS_r})~.
\end{equation}

The commutation relations (\ref{1.1}),(\ref{1.2}) in this
particular case acquire the form
\begin{eqnarray}
&&\lbrack
S_{lj},S_{lk}]=ie_{jkm}S_{lm},\;\;\;\;\;\;[R_{lj},R_{lk}]=ie_{jkm}S_{lm},
\nonumber\\
&&\lbrack \tilde{R}_{j},S_{lk}]=ie_{jkm}\tilde{R}_{m},\ \ \ \ \;\;
[\tilde{R}_{j},\tilde{R}_{k}] = ie_{jkm}S_{lm}, \nonumber \\
&&\lbrack R_{lj},S_{lk}]=ie_{jkm}R_{lm},\;\;\;\;\;
[R_{lj},\tilde{R}_{k}]=i\delta_{jk}A, \label{comm2} \\
&&\lbrack
\tilde{R}_{j},A] = iR_{lj}, \;\;\;\;[A,R_{lj}]=i\tilde{R}_{j},\;%
\;\;\;[A,S_{lj}]=0. \nonumber
\end{eqnarray}
The operators ${\bf R}_l$ and ${\bf \tilde{R}} $ are orthogonal to
${\bf S}_l$ in accordance with (\ref{orth}). Besides, ${\bf
R}_l\cdot {\bf\tilde{R}}=3X^{S_l S_r},$ and the Casimir operator
is ${\cal K}={\bf S}_l^{2}+{\bf R}_l^{2}+\tilde{\bf R}^{2}+A^2=4.$

Like in the case of double $SO(4)$ symmetry studied above, the
second step of RG procedure generates additional vertices in the
exchange part of the interaction Hamiltonian (\ref{sym5-tlr}),
\begin{eqnarray}
H_{cot}&=&J_{1}{\bf S}_{l}\cdot {\bf s}_l+J_{2}{\bf R}_{l}\cdot
{\bf s}_l  + J_3(\tilde{\bf R}_{1}\cdot {\bf s}_{rl}+\tilde{\bf
R}_{2}\cdot {\bf s}_{lr})\nonumber\\
&+&J_{4}{\bf S}_{l}\cdot {\bf s}_r+J_5{\bf \tilde{R}}\cdot {\bf
s}_l+J_6({\bf R}_{1l}\cdot {\bf s}_{rl}+ {\bf R}_{2l}\cdot {\bf
s}_{lr}) \nonumber\\
 &+&J_7{\bf S}_{l}\cdot({\bf s}_{lr}+{\bf s}_{rl})+
 J_8(\tilde{\bf R}_{1}\cdot {\bf s}_{lr}+\tilde{\bf
R}_{2}\cdot {\bf s}_{rl})\nonumber\\
&+&J_{9}{\bf R}_{l}\cdot {\bf s}_r+J_{10}{\bf \tilde{R}}\cdot {\bf
s}_r+J_{11}{\bf R}_{l}\cdot({\bf s}_{lr}+{\bf s}_{rl})\nonumber\\
&+&J_{12}({\bf R}_{1l}\cdot {\bf s}_{lr}+{\bf R}_{2l}\cdot {\bf
s}_{rl}),\label{sym52-tlr}
\end{eqnarray}
where ${\bf R}_{1l}=X^{S_lS_r}\tilde{\bf R}_{1}$, ${\bf
R}_{2l}=\tilde{\bf R}_{2}X^{S_rS_l}$. The scaling properties of
the system are determined by a system of 12 scaling equations with
initial conditions
\begin{eqnarray}
&&J_1(\bar D)=J^T_{l},\ \ \ \ \ \ \ J_2(\bar D)=J^{ST}_{l},\label{bcond5}\\
&&J_3(\bar D)=\alpha_r J_{lr}, \ \ \ \;J_{i}(\bar D)=0~~~~(i=4-12)
\nonumber
 \end{eqnarray}
(see Eq. \ref{J} for definitions) specifically
\begin{eqnarray}
\frac{dj_{1}}{d\ln d}&=&-\left[j_1^2 + j_2^2
+j_5^2+j_7^2+j_{11}^2+j_{11}(j_6+j_{12})\right.\nonumber\\
&+&\left.\frac{j_3^2+j_6^2+j_8^2+j_{12}^2}{2}\right],\nonumber\\
\frac{dj_{2}}{d\ln d} &=& -[2(j_1 j_2+j_7j_{11})+j_7(j_6+j_{12})-m_{lr}j_5],\nonumber\\
\frac{dj_{3}}{d\ln d} &=& -[j_3(j_1+j_4)+j_7(j_5+j_{10})-m_{lr}(j_6+j_{11})],\nonumber\\
\frac{dj_{4}}{d\ln d} &=& -\left[j_4^2+j_7^2+j_9^2+j_{10}^2+j_{11}^2+j_{11}(j_6+j_{12})\right.\nonumber\\
&+&\left.\frac{j_3^2+j_6^2+j_8^2+j_{12}^2}{2}\right],\nonumber\\
\frac{dj_{5}}{d\ln d} &=& -[2j_1j_5+j_7(j_3+j_{8})-m_{lr}j_2],\nonumber\\
\frac{dj_{6}}{d\ln d} &=& -[j_6(j_1+j_{4})-m_{lr}j_3],\nonumber\\
\frac{dj_{7}}{d\ln d} &=& -\left[\frac{(j_3
+j_8)(j_5+j_{10})+(j_2+j_9)(j_6+j_{12})}{2}\right.\nonumber\\
&+&\left.j_7(j_{1}+j_4)+j_{11}(j_2+j_{9})\right],\nonumber\\
\frac{dj_{8}}{d\ln d} &=& -[j_8(j_1+j_4)+j_7(j_5+j_{10})-m_{lr}(j_{11}+j_{12})],\nonumber\\
\frac{dj_{9}}{d\ln d} &=& -[2(j_4 j_9+j_7j_{11})+j_7(j_6+j_{12})-m_{lr}j_{10}],\nonumber\\
\frac{dj_{10}}{d\ln d} &=& -[2j_4j_{10}+j_7(j_3+j_{8})-m_{lr}j_9],\nonumber\\
\frac{dj_{11}}{d\ln d} &=& -[j_{11}(j_1+j_4)+j_7(j_2+j_{9})],\nonumber\\
\frac{dj_{12}}{d\ln d} &=& -[j_{12}(j_1+j_{4})-m_{lr}j_8].
 \label{lr-sc-eq5}
\end{eqnarray}
From equations (\ref{lr-sc-eq5}), one deduces the Kondo
temperature,
\begin{eqnarray}
T_{K2}=\bar{D}\Big(1-\frac{2\sqrt{2}m_{lr}}{j_1+j_2+\sqrt{(j_1+j_2)^2+2j_3^2}}\Big)
^{\displaystyle{\frac{1}{\sqrt{2}m_{lr}}}}. \label{T5-lr}
\end{eqnarray}
Similarly to the previous case, this equation transforms into the
usual exponential form when the $l-r$ symmetry of the leads is
perfect,
\begin{eqnarray}\label{tko1}
\lim_{m_{lr}\to 0}T_{K2} ={\bar D} e^{\displaystyle{
-\frac{2}{j_1+ j_2+\sqrt{(j_1+j_2)^2+2j_3^2}}}}.
\end{eqnarray}
Upon increasing $m_{lr}$, the symmetry reduces from $SO(5)$ to
$SO(4)$. The same happens at small $m_{lr}$ but with increasing
 $\bar \delta_{l}={\bar E}_{S_l}-{\bar E}_{T_l}$. In the latter case the
energy $\bar E_{S_l}$ is quenched, and at $\bar \delta_l \gg
T_{K2}$ Eq. (\ref{T5-lr}) transforms into $ T_{K}=\bar \delta_l
\exp\{-[j_1(\bar\delta_l)+ j_3(\bar\delta_l)]^{-1}\} $ (cf.
\cite{KKA02}). On the other hand, upon decreasing $\bar
\delta_{r}={\bar E}_{T_r}-{\bar E}_{S_l}$ the symmetry ${P}\times
SO(4)\times SO(4)$ is restored at $\bar \delta_{r}<T_{K0}$. The
Kondo effect disappears when $\bar \delta_{l}$ changes sign (the
ground state becomes singlet).

The next asymmetric configuration is illustrated by the flow
diagram of Fig.\ref{SO7}.
\begin{figure}[h]
\centerline{\epsfig{figure=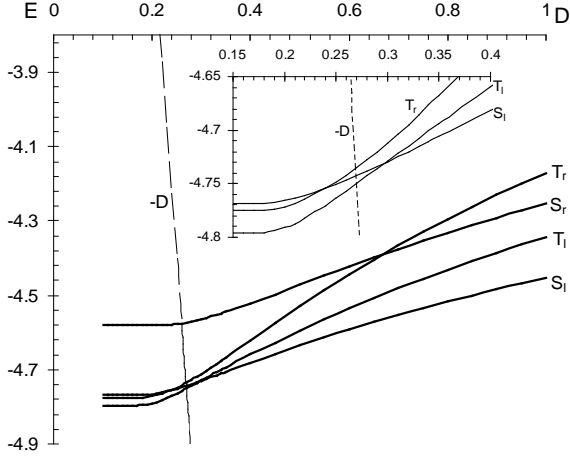,width=75mm,height=60mm,angle=0}}
\caption{Scaling trajectories for $ SO(7)$ symmetry in the SW
regime. Inset: Zoomed in avoided level crossing pattern near the SW line}
\label{SO7}
\end{figure}

In this case, the manifold $\{T_l,S_l,T_r\}$ is involved in the
dynamical symmetry of TQD. The relevant symmetry group is $SO(7)$.
It is generated by six vectors and three scalars. These are spin
operators ${\bf S}_a$ $(a=l,r)$ and R-operator ${\bf R}_l$  (see
Eq. \ref{comm1}) plus three vector operators ${\bf \tilde{R}}_i$
and three scalar operators $A_i$ involving $l-r$ permutation. Here
are the expressions for the spherical components of these vectors
via Hubbard operators,
\begin{eqnarray}
&&\tilde{R}_1^{+} =\sqrt{2}( X^{1_r0_l}+X^{0_l\bar{1}_r}),\;\;\;
\tilde{R}_1^{z} =X^{1_l1_r}-X^{{\bar 1}_r{\bar 1}_l},\nonumber\\
&&\tilde{R}_2^{+}=\sqrt{2}( X^{1_l0_r}+X^{0_r\bar{1}_l}),\ \
\tilde{R}_2^{z} =X^{1_r1_l}-X^{{\bar 1}_l{\bar 1}_r}, \label{so7}\\
&&\tilde{R}^{+}_3 =\sqrt{2}( X^{1_rS_l}-X^{S_l\bar{1}_r}),\ \;
\tilde{R}_3^{z} =-(X^{0_rS_l}+X^{S_l0_r}).\nonumber
\end{eqnarray}
The scalar operators $A_1$, $A_2$, $A_3$ now involve the $l-r$
permutations for the triplet states. They are defined as
\begin{eqnarray}
A_1 &=&\frac{i\sqrt{2}}{2}\left(X^{1_r{\bar 1}_l}-X^{1_l{\bar
1}_r+
X^{{\bar 1}_r1_l}-X^{{\bar 1}_l1_r}}\right ),\nonumber\\
A_2 &=&\frac{\sqrt{2}}{2}\left(X^{1_l{\bar 1}_r}-X^{1_r{\bar
1}_l}+X^{{\bar 1}_r1_l}-
 X^{{\bar 1}_l1_r}\right),\nonumber\\
A_3&=&i\left(X^{0_l0_r}-X^{0_r0_l}\right).\label{A}
\end{eqnarray}
The (somewhat involved) commutation relations of $o_7$ algebra for
these operators and various kinematic constraints are presented in
Appendix \ref{algebra-o7}. The SW transformation results in the
effective cotunneling Hamiltonian
\begin{eqnarray}
H_{cot}&=&\sum_{\Lambda=T_l,S_l,T_r}{\bar{E}}_{\Lambda}X^{\Lambda\Lambda}
+M_{lr}(X^{T_lT_r}+X^{T_rT_l})\nonumber\\
&+&\sum_{ k \sigma}\sum_{a=l,r}\epsilon_{ka}
c^{+}_{ak\sigma}c_{ak\sigma}+t_{lr}\sum_{ k \sigma}\sum_{a=l,r}
c^{+}_{ak\sigma}c_{{\bar a}k\sigma }
\nonumber\\
&+&\sum_{a=l,r} J_{1a}{\bf S}_{a}\cdot {{\bf s}}_a+
J_2\sum_{a=l,r}{\bf
{S}}_{a{\bar a}}\cdot {\bf s}_{a{\bar a}}\nonumber\\
&+&J_{3}({\bf {\tilde R}}^{(1)}_{3}\cdot {\bf s}_{rl}+ {\bf
{\tilde R}}^{(2)}_{3}\cdot {\bf s}_{lr})+J_4{\bf {R}}_{l}\cdot
{{\bf s}_l}, \label{int7}
\end{eqnarray}
where $J_{1a}=J_a^T$, $J_{2}=J_{{lr}}$, $J_{3}=\alpha_l J_{lr}$,
$J_{4}=\alpha_l J_{l}^T$ and ${\bf {S}}_{a{\bar
a}}=\sum_{\mu}X^{\mu_a\mu_{\bar a}}{\bf {S}}_{{\bar a}}$. The
spherical components of the vector operators $\tilde{\bf R}_1$ and
$\tilde{\bf R}_2$ are
\begin{eqnarray}
\tilde{R}^{(1)+}_3 &=&\sqrt{2}X^{1_rS_l},\ \ \ \;
\tilde{R}^{(1)-}_3=-\sqrt{2} X^{{\bar 1}_rS_l},\nonumber\\
\tilde{R}^{(2)+}_3&=&(\tilde{R}^{(1)-}_3)^\dag,\ \ \ \
\tilde{R}^{(2)-}_3=(\tilde{R}^{(1)+}_3)^\dag,\nonumber\\
\tilde{R}^{(1)}_{3z} &=&-X^{0_rS_l},\ \ \ \ \;
\tilde{R}^{(2)}_{3z}=(\tilde{R}^{(1)}_{3z})^\dag.
\end{eqnarray}
It is easy to see that ${\bf S}_{lr}+{\bf S}_{rl}={\bf {\tilde
R}}_1+{\bf {\tilde R}}_2$ and ${\bf {\tilde R}}_3={\bf {\tilde
R}}^{(1)}_{3}+{\bf {\tilde R}}^{(2)}_{3}$.

Like in the case of $SO(5)$ symmetry, the tunneling terms
$M_{lr}X^{T_a T_{\bar a}}$ generate additional vertices in the
renormalized Hamiltonian $H_{cot}$. The number of these vertices
and the corresponding scaling equations is too wide to be
presented here. We leave the description of RG procedure for
$SO(7)$ group for the next section (as well as the case of TQD
with odd occupation), where the case of $M_{lr}=0$ is considered.
In that case the scaling equations describing the Kondo physics of
TQD with $SO(n)$ symmetry are more compact.
\subsection{Section summary}
The basic physics for all $SO(n)$ symmetries is the same, and we
summarize it here. We have analyzed several examples of TQD with
even occupation in the parallel geometry (Fig. \ref{TQD}). Our
analysis demonstrates the principal features of Kondo effect in
CQD in comparison with the conventional SQD composed of a single
well. These examples teach us that in Kondo tunneling through CQD,
not only the spin rotation but also the ''Runge-Lenz'' type
operators ${\bf R}$ and $\tilde{\bf R}$ are involved. Physically,
the operators $\tilde{\bf R}$ describe left-right transitions, and
different Young schemes give different spin operators in the
effective co-tunneling Hamiltonians (see Appendix \ref{tab}).

\section{Triple quantum dot in series} \label{IV}
\subsection{Motivation}\label{IV A}
It was mentioned already in Section \ref{III} that a TQD with
leads possessing perfect $l-r$ symmetry can be mapped onto a TQD
in a series by means of geometrical conformal transformation.
Indeed, if the inter-channel tunneling amplitude $t_{lr}$ in the
Hamiltonian (\ref{tlead}) is zero, one may apply the rotation in
the source-drain space separately to each channel and exclude the
odd $s-d$ combination of lead states both in the $l$- and
$r$-channel. \cite{Glaz} Since now each lead is coupled to its own
reservoir, and one arrives at the series configuration shown in
Fig.\ref{TQD-s}.

It is virtually impossible to conceive an additional
transformation after which  the odd combination of lead states are
excluded from the tunneling Hamiltonian \cite{EPL}. As a result,
the challenging situation arises in case of odd occupation $N=3$,
where the net spin of TQD is $S=1/2$, and the two leads play part
of two channels in Kondo tunneling Hamiltonian. Unfortunately,
despite the occurrence of two electron channels in the spin
Hamiltonian, the complete mapping on the two-channel Kondo problem
is not attained because there is an additional cotunneling term
$J_{lr} {\bf S}\cdot {\bf s}_{lr}+H.c.$ (${\bf s}_{lr}$ is
determined by (\ref{1.100})) which turns out to be relevant, and
the two-channel fixed point cannot be reached. And yet, from the
point of view of dynamical symmetry the series geometry offers a
new perspective which we analyze in the present section for cases
of even and odd occupation.

\subsection{Even occupation}\label{IV B}
\begin{figure}[htb]
\centering
\includegraphics[width=75mm,height=70mm,angle=0]{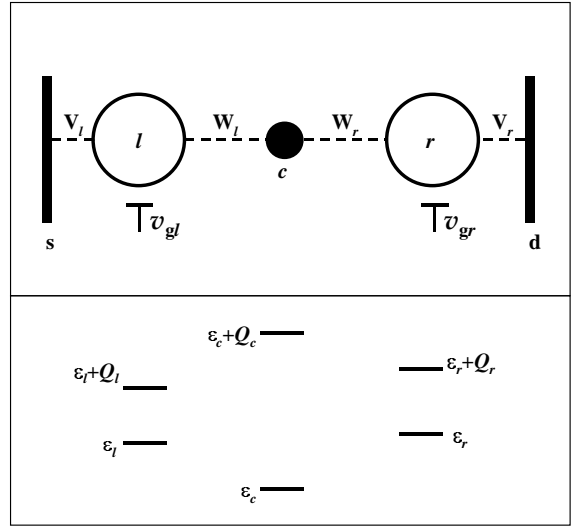}
\caption{Triple quantum dot in series. Left $(l)$ and right $(r)$
dots are coupled by tunneling $W_{l,r}$ to the central $(c)$ dot
and by tunneling $V_{l,r}$ to the source $(s)$ (left) and drain
$(d)$ (right) leads.}\label{TQD-s}
\end{figure}

Consider then a TQD in series (Fig.\ref{TQD-s}) with four electron
occupation $N=4$. The Hamiltonian of the system can be written in
the form,
\begin{eqnarray}
H&=&\sum_{\Lambda_{a}}{E}_{\Lambda_a} X^{\Lambda_a\Lambda_a}
+\sum_{\lambda}{E}_\lambda X^{\lambda\lambda}
 +\sum_{
k \sigma}\sum_{b=s,d}\epsilon_{kb} c^{+}_{bk\sigma}c_{bk\sigma}
\nonumber\\
& +&
\sum_{\Lambda\lambda}\sum_{k\sigma}[(V^{\lambda\Lambda}_{l\sigma}
c^{+}_{sk\sigma }+V^{\lambda\Lambda}_{r\sigma}
c^{+}_{dk\sigma})X^{\lambda\Lambda}+ H.c.]. \label{Hser}
\end{eqnarray}
 Here $|\Lambda\rangle$, $|\lambda\rangle$ are the four- and
 three-electron eigenfunctions (\ref{eg-fun4}) and (\ref{eg-func}), respectively;
 ${E}_\Lambda,$ ${E}_\lambda$ are
 the four- and three-electron energy levels,
 respectively; $X^{\lambda\Lambda}=|\lambda\rangle\langle\Lambda|$ are
 number changing dot Hubbard
 operators. The tunneling amplitudes
 $V^{\lambda\Lambda}_{a\sigma}=
V_a\langle\lambda|d_{a\sigma}|\Lambda\rangle$ ($a=l,r$) depend
explicitly on the respective $3-4$ particle quantum numbers
$\lambda$, $\Lambda$. Note that direct tunneling through the TQD
is suppressed due to electron level mismatch and Coulomb blockade,
so that only {\it cotunneling} mechanism contributes to the
current.

The generic Hamiltonian (\ref{generic}) simplifies in this case to
\begin{eqnarray}
H_{cot}&=&\sum_{\Lambda_a}{\bar
E}_{\Lambda_a}X^{\Lambda_a\Lambda_a}
+\sum_{k\sigma}\sum_{b=l,r}\epsilon_{kb}
c^{\dagger}_{bk\sigma}c_{bk\sigma}\nonumber\\
&+&
 \sum_{a=l,r}J^T_{a} {\bf S}_{a}\cdot {\bf s}_a+ J_{lr}{\hat P}
\sum_{a=l,r}{\bf S}_{a}\cdot {\bf s}_{{\bar a}a}\nonumber\\
&+&\sum_{a=l,r}J^{ST}_a {\bf R}_{a}\cdot {\bf s}_a
+J_{lr}\sum_{a=l,r}\tilde{\bf R}_{a}\cdot {\bf s}_{a{\bar a}},
\label{ex-H}
\end{eqnarray}
(the notation $l,r$ is used for the electron states both in the
leads and in the TQD). The antiferromagnetic coupling constants
are defined by (\ref{J}). The vectors ${\bf S}_a$, ${\bf R}_a$ and
$\tilde{\bf R}_a$ are the dot operators (\ref{comm1}), $\hat P$ is
the permutation operator (\ref{perm}), and the components of the
vectors ${\bf s}_a$, ${\bf s}_{a\bar a}$ are determined in Eqs.
(\ref{1.10}) (with $\alpha=a=l,r$) and (\ref{1.100}). The vector
operators ${\bf S}_{a},$ ${\bf R}_{a}$, $\tilde{\bf R}_{a}$ and
the permutation operator ${\hat P}$ manifest the dynamical
symmetry of the TQD.

We now discuss possible realization of $P\times SO(4)\times
SO(4)$, $SO(5)$ and $SO(7)$ symmetries arising in the TQD with
$N=4$. Due to the absence of interchannel mixing, the avoided
crossing effect does not arise in the series geometry. As a
result, all cases of high symmetry are characterized by the same
flow diagram of Figs. \ref{pdotso4},\ref{SO5} and \ref{SO7} but
{\it without avoided crossing effects shown in the insets.}

Let us commence the analysis of
the Kondo effect in the series geometry with the
case ${P}\times SO(4)\times SO(4)$ where $E_{lc}=E_{rc}$ and
${\Gamma_{T_r}}={\Gamma_{T_l}}$ (Fig.\ref{pdotso4}). In this case
the exchange part of the Hamiltonian (\ref{ex-H}) is a simplified
version of the Hamiltonian (\ref{sym-h-mlr}) with the boundary
conditions (\ref{bcond}). The scaling equations are the same as
(\ref{sc-eq-so4}) with $m_{lr}=0$. Solving them one gets
Eq.(\ref{tko}) for the Kondo temperature.

When ${\bar E}_{S_l}\approx {\bar E}_{T_l}\approx {\bar E}_{S_r}<
{\bar E}_{T_r}$ (Fig.\ref{SO5}), the TQD possesses the $SO(5)$
symmetry. In this case the interaction Hamiltonian has the form
\begin{equation}
H_{cot}=J_{1}{\bf S}_{l}\cdot {\bf s}_l+J_{2}{\bf R}_{l}\cdot {\bf
s}_l+ J_3(\tilde{\bf R}_{1}\cdot {\bf s}_{rl}+\tilde{\bf
R}_{2}\cdot {\bf s}_{lr}), \label{sym5-ser}
\end{equation}
which is the same as in Eq.(\ref{sym5-tlr}) with $\tilde{\bf
R}_{1},\ \tilde{\bf R}_{2}$ determined by Eq.(\ref{r-tild}).
Respectively, the effective Hamiltonian for the Anderson scaling
is a reduced version of the Hamiltonian (\ref{sym52-tlr})
\begin{eqnarray}
H_{cot}&=&J_{1}{\bf S}_{l}\cdot {\bf s}_l+J_{2}{\bf R}_{l}\cdot
{\bf s}_l\nonumber\\
&+& J_3(\tilde{\bf R}_{1}\cdot {\bf s}_{rl}+\tilde{\bf R}_{2}\cdot
{\bf s}_{lr})+J_{4}{\bf S}_{l}\cdot {\bf s}_r, \label{sym52}
\end{eqnarray}
with the boundary conditions (\ref{bcond5}) for $J_i, ~i=1-4$.

The scaling equations have the form
\begin{eqnarray}
\frac{dj_{1}}{d\ln d} &=& -\left[j_1^2 + j_2^2
+\frac{j_3^2}{2}\right],\nonumber\\
\frac{dj_{2}}{d\ln d} &=& -2j_1 j_2,\nonumber\\
\frac{dj_{3}}{d\ln d} &=& -j_3(j_1+j_4),\nonumber\\
\frac{dj_{4}}{d\ln d} &=& -\left[j_4^2
+\frac{j_3^2}{2}\right].\label{ser-sc-eq55}
\end{eqnarray}
Of course, Eqs.(\ref{ser-sc-eq55}) for the Kondo temperature yield
the limiting value (\ref{tko1}).

 When ${\bar E}_{T_l}\approx {\bar E}_{T_r}< {\bar
E}_{S_l},{\bar E}_{S_r}$ (Fig.\ref{SO7}), the TQD possesses the
$SO(7)$ symmetry. In this case the Anderson RG procedure adds
three additional vertices in the exchange part of the basic SW
Hamiltonian (\ref{int7}),
\begin{eqnarray}
H_{cot}&=&\sum_{a=l,r} J_{1a}{\bf S}_{a}\cdot {{\bf s}}_a+
J_2\sum_{a=l,r}{\bf
{S}}_{a{\bar a}}\cdot {\bf s}_{a{\bar a}}\nonumber\\
&+&J_{3}({\bf {\tilde R}}^{(1)}_{3}\cdot {\bf s}_{rl}+ {\bf
{\tilde R}}^{(2)}_{3}\cdot {\bf s}_{lr})+J_4{\bf {R}}_{l}\cdot
{{\bf s}_l}\nonumber\\
&+&\sum_{a=l,r} J_{5a}{\bf S}_{a}\cdot {{\bf s}}_{\bar a}+J_6{\bf
{R}}_{l}\cdot {{\bf s}_r}. \label{int7-1}
\end{eqnarray}
The boundary conditions for solving the scaling equations are
\begin{eqnarray}
&& J_{1a}(\bar D)=J^T_a,~~J_2(\bar D)=J_{lr}, ~~J_3(\bar
D)=\alpha_l J_{lr},\label{bound7}\\
&&J_4(\bar D)=\alpha_l J_{l}^T, ~~J_{5a}(\bar D)=J_6(\bar D)=0~
(a=l,r).\nonumber
\end{eqnarray}

The system of scaling equations
\begin{eqnarray}
\frac{dj_{1l}}{d\ln d}&=&-\left[j_{1l}^2 + \frac{j_2^2}{2}
+j_4^2\right],\nonumber\\
\frac{dj_{1r}}{d\ln d}&=&-\left[j_{1r}^2 + \frac{j_2^2}{2}
+\frac{j_3^2}{2}\right],\nonumber\\
\frac{dj_{2}}{d\ln d} &=& -\frac{j_2(j_{1l}+j_{1r}+j_{5l}+j_{5r})+ j_3(j_4+j_6)}{2},\nonumber\\
\frac{dj_{3}}{d\ln d}&=&-\left[j_2(j_4+j_6)+j_3(j_{1r}+j_{5r})\right],\nonumber\\
\frac{dj_{4}}{d\ln d} &=& -\left[2j_{1l}j_4+j_2j_3\right],\nonumber\\
\frac{dj_{5l}}{d\ln d}&=&-\left[j_{5l}^2 + \frac{j_2^2}{2}
+j_6^2\right],\nonumber\\
\frac{dj_{5r}}{d\ln d}&=&-\left[j_{5r}^2 + \frac{j_2^2}{2}
+\frac{j_3^2}{2}\right],\nonumber\\
\frac{dj_{6}}{d\ln d} &=& -\left[2j_{5l}j_6+j_2j_3\right]
\label{ser-sc-eq7}
\end{eqnarray}
is now solvable analytically, and the Kondo temperature is,
\begin{eqnarray}
T_{K}=\bar{D}\exp\left\{\frac{4}{2j_{+}+\sqrt{4j_{-}^2+3(j_2+j_3)^2}}\right\}
, \label{T7-ser}
\end{eqnarray}
where $j_+=j_{1l}+j_4+j_{1r}$, $j_-=j_{1l}+j_4-j_{1r}$.

Like in the cases considered above, the Kondo temperature and the
dynamical symmetry itself depend on the level splitting. On
quenching the $S_l$ state (increasing $\bar\delta_{lr}={\bar
E}_{S_l}-{\bar E}_{T_r}$), the pattern is changed into a
${P}\times SO(3)\times SO(3)$ symmetry of two degenerate triplets
with a mirror reflection axis. Changing the sign of $\delta_{lr}$
one arrives at a singlet regime with $T_K=0$.
\begin{figure}[h]
\centerline{\epsfig{figure=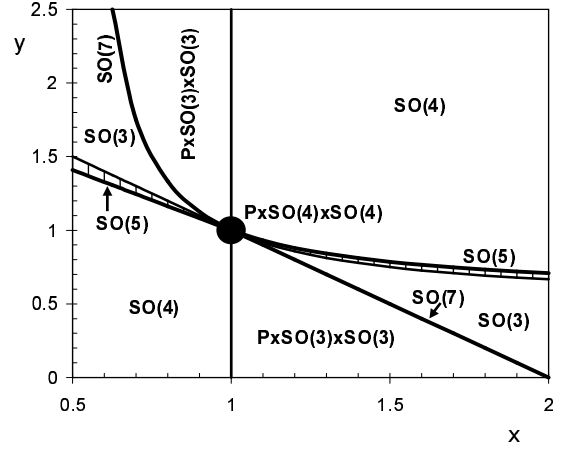,width=72mm,height=58.8mm,angle=0}}
\caption{Phase diagram of TQD. The numerous dynamical symmetries
of a TQD in the parallel geometry are presented in the plane of
experimentally tunable  parameters $x=\Gamma_l/\Gamma_r$ and
$y=E_{lc}/E_{rc}.$}\label{phasediag}
\end{figure}

The results of calculations described in this subsection are
summarized in Fig.\ref{phasediag}. The central domain of size
$T_{K0}$ describes the fully symmetric state where there is
left-right symmetry. Other regimes of Kondo tunneling correspond
to lines or segments in the $\{x,y\}$ plane. These lines
correspond to cases of higher conductance (ZBA). On the other
hand, at some regions, the TQD has a singlet ground state and the
Kondo effect is absent. These are marked by the vertically hatched
domain. Both the tunneling rates which enter the ratio $x$ and the
relative level positions which determine the parameter $y$ depend
on the applied potentials, so the phase diagram presented in
Fig.\ref{phasediag} can be scanned {\it experimentally} by
appropriate variations of $V_a$ and $v_{ga}.$ This is a rare
occasion where an abstract concept like dynamical symmetry can be
felt and tuned by experimentalists. The quantity that is measured
in tunneling experiments is the zero-bias anomaly (ZBA) in tunnel
conductance $g$. \cite{KKK} The ZBA peak is strongly temperature
dependent, and this dependence is scaled by $T_K$. In particular,
in a high temperature region $T>T_K$, where the scaling approach
is valid, the conductance behaves as
\begin{equation}\label{cond}
g(T) \sim \ln^{-2}(T/T_K).
\end{equation}
As it has been demonstrated above, $T_K$ in CQD is a non-universal
quantity due to partial break-down of dynamical symmetry in these
quantum dots. It has a maximum value in the point of highest
symmetry ${P}\times SO(4)\times SO(4)$, and depends on the
parameters $\delta_a$ in the less symmetric phases (see, e.g.,
Eqs. \ref{tko}, \ref{Tdelta}, \ref{tko1}, \ref{T7-ser}). Thus,
scanning the phase diagram means changing $T_K(\delta_a)$.
\begin{figure}[h]
\centerline{\epsfig{figure=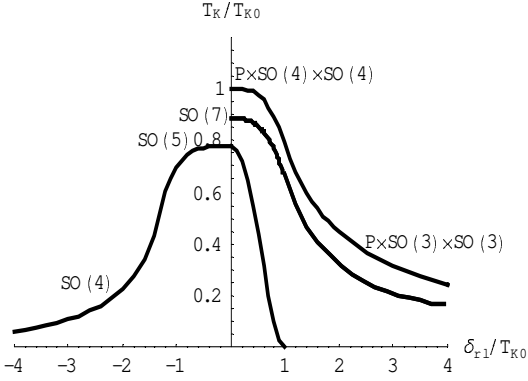,width=70mm,height=50mm,angle=0}}
\caption{Variation of Kondo temperature with $\delta_{rl}\equiv
v_{gr}-v_{gl}$. Increasing this parameter removes some of the
degeneracy and either "breaks" or reduces the corresponding
dynamical symmetry.} \label{vartek}
\end{figure}
These changes are shown in Fig.\ref{vartek} which illustrates the
evolution of $T_K$ with $\delta_{rl}$ for $x=0.96$, $0.8$ and 0.7
corresponding to a symmetry change from ${P}\times SO(4)\times
SO(4)$, $SO(7)$ to ${P}\times SO(3)\times SO(3)$ and from $SO(5)$
to $SO(4)$, respectively. It is clear that the conductance
measured at given $T$ should follow variation of $T_K$ in
accordance with (\ref{cond}).

\subsection{Odd occupation}\label{IV C}
We now turn our attention to investigation of the dynamical
symmetries of TQD in series with odd occupation $N=3$, whose
low-energy spin multiplet contains two spin 1/2 doublets
$|B_{1,2}\rangle$ and a spin quartet $|Q\rangle$,
\begin{eqnarray}
E_{B_1}&=&{\varepsilon}_c +{\varepsilon}_l +{\varepsilon}_r- \frac
{3}{2}\left[W_l\beta_l+W_r\beta_r \right], \nonumber \\
E_{B_2}&=&{\varepsilon}_c +{\varepsilon}_l +{\varepsilon}_r-
\frac{1}{2}\left[W_l\beta_l+W_r\beta_r\right], \nonumber \\
E_{Q}&=&{\varepsilon}_c +{\varepsilon}_l +\varepsilon_r.
\label{Edot}
\end{eqnarray}
There are also four charge-transfer excitonic counterparts of the
spin doublets separated by the charge transfer gaps $\sim
\varepsilon_l-\varepsilon_c +Q_l$ and $\varepsilon_r-\varepsilon_c
+Q_r$ from the above states (see Appendix \ref{diag}).

\begin{figure}[h]
\centerline{\epsfig{figure=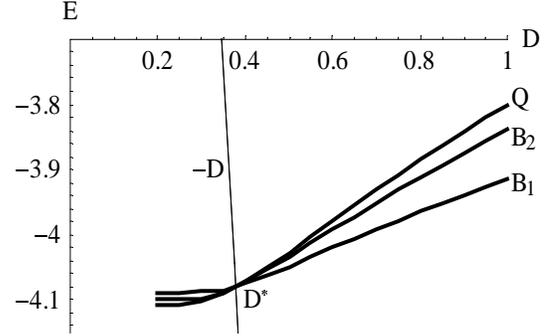,width=70mm,height=45mm,angle=0}}
\caption{Scaling trajectories resulting in $SO(4)\times SU(2)$
symmetry of TQD with $N=3$.}\label{crossym}
\end{figure}

Like in the four-electron case, the scaling equations (\ref{DE2})
may be derived with different tunneling rates for different spin
states ($\Gamma_{Q}$ for the quartet and $\Gamma_{B_{i}} (i=1,2)$
for the doublets).
\begin{eqnarray}
&&\Gamma_{Q} = \pi\rho_0 \left(V_{l}^2+ V_{r}^2\right), \nonumber \\
&&\Gamma_{B_1} = \gamma_1^2\Gamma_{Q}, \ \ \ \Gamma_{B_2} =
\gamma_2^2\Gamma_{Q},\label{Gamma}
\end{eqnarray}
with
\begin{equation}
\gamma_1=\sqrt{1-\frac{3}{2}\left(\beta_l^2+\beta_r^2\right)},~~
\gamma_2=\sqrt{1-\frac{1}{2}\left(\beta_l^2+\beta_r^2\right)}.\label{gam}
\end{equation}
Since $\Gamma_{Q}>\Gamma_{B_1},\Gamma_{B_2 }$, the scaling
trajectories cross in a unique manner: This is the complete
degenerate configuration where all three phase trajectories
$E_{\Lambda}$ intersect
[$E_{Q}(D^{\star})=E_{B1}(D^{\star})=E_{B2}(D^{\star})$] at the
same point $D^{\star}.$ This happens at bandwidth $D=D^{\star}$
(Fig.\ref{crossym}) whose value is estimated as
\begin{eqnarray}
D^{\star} &=& D_0\exp \left(-\frac{\pi
r}{\Gamma_{Q}}\right),\label{Dc}
\end{eqnarray}
where
$$r=\frac{W_l^2E_{rc}+W_r^2E_{lc}}{W_l^2E_{rc}^2+W_r^2E_{lc}^2}E_{lc}E_{rc}.$$
If this degenerate point occurs in the SW crossover region, i.e.,
if $D^{\star}\approx \bar{D}$, the SW procedure involves all three
spin states, and it results in the following cotunneling
Hamiltonian
\begin{eqnarray}
&&H_{cot}=\sum_{a=l,r}(J_{a}^{T}{\bf S} +J_{a}^{ST} {\bf R}
 )\cdot {\bf s}_a , \label{HSO4}
\end{eqnarray}
where ${\bf S}$ is the spin 1 operator and ${\bf R}$ is the
R-operator describing S/T transition similar to that for spin
rotator \cite{KA02}. The coupling constants are
\begin{eqnarray}
J_{a}^{T}= \frac{4 \gamma_1
|V_{a}|^{2}}{3(\epsilon_{F}-\varepsilon_{a})},\ \ \
J_{a}^{ST}=\gamma_2 J_{lr}^{T}.
\end{eqnarray}

This is a somewhat unexpected situation where Kondo tunneling in a
quantum dot with {\it odd} occupation demonstrates the exchange
Hamiltonian of a quantum dot with {\it even} occupation. The
reason for this scenario is the specific structure of the wave
function of TQD with $N=3$. The corresponding wave functions
$|\Lambda\rangle$ (see Appendix \ref{diag}) are vector sums of
states composed of a "passive" electron sitting in the central dot
and singlet/triplet (S/T) two-electron states in the $l,r$ dots.
Constructing the eigenstates $|\Lambda \rangle$ using certain
Young tableaux (see Appendix D), one concludes that the spin
dynamics of such TQD is represented by the spin 1 operator ${\bf
S}$ corresponding to the $l-r$ triplet, the corresponding
R-operator ${\bf R}$ and the spin 1/2 operator ${\bf s}_c$ of a
passive electron in the central well. The latter does not enter
the effective Hamiltonian $H_{cot}$ (\ref{HSO4}) but influences
the kinematic constraint via Casimir operator ${\cal K}={\bf
S}^{2}+{\bf M}^{2}+{\bf s}^{2}_c=\frac {15} {4}$. The dynamical
symmetry is therefore $SO(4) \times SU(2)$, and only the $SO(4)$
subgroup is involved in Kondo tunneling.

The scaling equations have the form,
\begin{eqnarray}
\frac{dj_{1a}}{d\ln d} &=&
-[j_{1a}^2+j_{2a}^2],\nonumber\\
\frac{dj_{2a}}{d\ln d} &=& -2j_{1a}j_{2a},
 \label{sc-42}
\end{eqnarray}
where $j_{1a}=\rho_0J_{a}^T$, $j_{2a}=\rho_0J_{a}^{ST}$ $(a=l,r)$.
From Eqs. (\ref{sc-42}) we obtain the Kondo temperature,
\begin{equation}
T_{K}=\rm{max}\{T_{Kl},T_{Kr}\}, \label{T3s}
\end{equation}
with $T_{Ka}={\bar D}\exp\left[-1/(j_{1a}+j_{2a})\right]$.

 An
additional dynamical symmetry arises in the case when $D^\star>
\bar{D}$. In this case the ground state of TQD is a quartet S=3/2,
and we arrive at a standard underscreened Kondo effect for $SU(2)$
quantum dot  as an ultimate limit of the above highly degenerate
state.

\subsection{Section summary}
 To conclude this section, it might be useful here to
underscore the following points: 1) The difference between series
and parallel geometries of TQD coupled to the leads by two
channels exists only at non-zero interchannel mixing in the leads,
$t_{lr}\neq 0$.  2) One may control the dynamical symmetry of
Kondo tunneling through TQD by varying the gate voltage and/or
lead-dot tunneling rate. 3) In the case of odd electron occupation
($N=3$) when the ground-state of the isolated TQD is a doublet and
higher spin excitations can be neglected, the effective low-energy
Hamiltonian of a TQD in series manifests a two-channel Kondo
problem albeit {\it only in the weak coupling regime}.\cite{EPL}
To describe the flow diagram in this case, one should go beyond
the one-loop approximation in RG flow equations. \cite{NB} 4) The
nominal spin of CQD does not necessarily coincide with that
involved in Kondo tunneling. A simple albeit striking realization
of this scenario in this context is the case of TQD with $N=3$,
which manifests itself as a dot with integer or half-integer spin
depending on gate voltages.
\section{Anisotropic Kondo tunneling through TQD in series
geometry} \label{V}
\subsection{Generalities}
In all examples of CQDs considered in the previous sections the
co-tunneling problem is mapped on the specific spin Hamiltonian
where both ${\bf S}$ and ${\bf R}$ vectors are involved in
resonance cotunneling. There are, however, more exotic situations
where the effective spin Hamiltonian is in fact a "Runge-Lenz"
Hamiltonian in the sense that the vectors ${\bf R}$ {\it alone}
are responsible for Kondo effect. Actually, just this aspect of
dynamical symmetry in Kondo tunneling was considered in the
theoretical papers cited in Ref. \onlinecite{Magn} and observed
experimentally in Ref. \onlinecite{ext}, in which the Kondo effect
in planar and vertical QDs induced by external magnetic field $B$
has been observed. In this section we lay down the theoretical
basis for this somewhat unusual kind of Kondo effect.

Consider again the case of TQD in series geometry with $N=4$. In
the previous sections the variation of spin symmetry is due to the
interplay of two contributions to indirect exchange coupling
between the spins ${\bf S}_a$. One source of such an exchange is
tunneling within the CQD (amplitudes $W_a$) and another one is the
tunneling between the dots and the leads (amplitudes $V_a$). An
appropriate tuning of these two contributions results in
occasional degeneracy of spin states (elimination of exchange
splitting), and various combinations of these occasional
degeneracies lead to the rich phase diagram presented in Fig.
\ref{phasediag}. A somewhat more crude approach, yet more
compliant with experimental observation of such interplay is
provided by the Zeeman effect. This mechanism is effective for CQD
which remains in a singlet ground state after all exchange
renormalizations have taken place. The negative exchange energy
$\delta_a$ may then be compensated by the Zeeman splitting of the
nearest triplet states, and Kondo effect arises once this
compensation is complete.\cite{Magn} From the point of view of
dynamical symmetry, the degeneracy induced by magnetic field means
realization of one possible subgroup of the non-compact group
$SO(n)$ (see Eq. \ref{1.14} and corresponding discussion in
Section \ref{II}). The transformation $SO(4)\to SU(2)$ for DQD in
magnetic field was discussed in Ref. \onlinecite{KA02}.

\subsection{Quantum dot with SU(3) dynamical symmetry}
In similarity with DQD, the Kondo tunneling may be induced by
external field $B$ in the non-magnetic sector of the phase diagram
of Fig. \ref{phasediag}. A very peculiar Kondo tunneling is
induced by an external magnetic field $B$ in the non-magnetic
sector of the phase diagram of Fig. \ref{phasediag} close to the
$SO(5)$ line. In this case, a remarkable symmetry reduction occurs
when the Zeeman splitting compensates negative
$\delta_{l,r}=E_{S_l,r}-E_{T_l}$.
 Then we are left in the subspace of states
$\{T1_l,S_l, S_r\}$, and the interaction Hamiltonian has the form,
\begin{eqnarray}
\widetilde{H}_{cot}
&=&(J_{1}R^z_{1}+J_{2}R^z_{2})s^z_l+\frac{\sqrt{2}}{2}J_{3l}
\left(R^{+}_{1}s^{-}_l+R^{-}_{1}s^{+}_l\right)\nonumber\\
&+&\frac{\sqrt{2}}{2}J_{3r}(R^{+}_{2}s^{-}_{lr}+R^{-}_{2}s^{+}_{rl})
+J_{4}\left(R_{3}s^{z}_{lr}+R_{4}s^{z}_{rl}\right)\nonumber\\
&+&(J_5R^z_1+J_6R^z_2)s^z_r
+J_7(R^{+}_{1}s^{-}_r+R^{-}_{1}s^{+}_r).\label{int7m-s}
\end{eqnarray}
Here
\begin{eqnarray}
 J_{1}(\bar D)&=&J_{2}(\bar D) =\frac{2J^{T}_{l}}{3},\ \ \ \ J_{3l}(\bar
D)=J^{ST}_{l},\nonumber\\
J_{3r}(\bar D)&=&\alpha_r J_{lr},\ \ \ \ \ J_i(\bar D)=0 \ \
(i=4-7).
\end{eqnarray}
The operators ${\bf R}_{1}$, ${\bf R}_{2}$, $R_{3}$ and $R_{4}$
are defined as,
\begin{eqnarray}
R^z_{1}&=&\frac{1}{2}(X^{1_l1_l}-X^{S_lS_l}),\ \;
R^{+}_{1}=X^{1_lS_l}, \ \;
R^{-}_{1}=(R^{+}_{1})^{\dag},\nonumber\\
R^z_{2}&=&\frac{1}{2}(X^{1_l1_l}-X^{S_rS_r}),\
R^{+}_{2}=X^{1_lS_r}, \ \;
R^{-}_{2}=(R^{+}_{2})^{\dag},\nonumber\\
R_{3}&=&\frac{\sqrt 3}{2}X^{S_lS_r},\ \ \ \ \ \ \ \ \ \ \ \ \ \ \
\ R_{4}=\frac{\sqrt 3}{2}X^{S_rS_l}. \label{RU3}
\end{eqnarray}
We see that the anisotropic Kondo Hamiltonian (\ref{int7m-s}) is
quite unconventional. There are several different terms
responsible for transverse and longitudinal exchange involving the
R-operators which generate both S$_a$/T and S$_a$/S$_{\bar a}$
transitions.

The operators (\ref{RU3}) obey the following commutation
relations,
\begin{eqnarray}
&&\lbrack R_{1j},R_{1k}]=ie_{jkm}R_{1m},\ \ \ \ \ \ \;\;
[R_{2j},R_{2k}]=i e_{jkm}R_{2m}, \nonumber\\
&&\lbrack R_{1j},R_{2k}]=
\frac{\sqrt{3}}{6}(R_3-R_4)\delta_{jk}(1-\delta_{jz})\nonumber\\
&&\ \ \ +\frac{i}{2}e_{jkm}\Big(R_{1m}\delta_{kz}+
R_{2m}\delta_{jz}-\frac{\sqrt{3}}{3}\delta_{mz}(R_3+R_4)\Big), \nonumber\\
&&\lbrack R_{1j},R_{3}]=-\frac{1}{2}R_3\delta_{jz}+
\frac{\sqrt{3}}{4}(R_{2x}+i R_{2y})(\delta_{jx}-i\delta_{jy}),\nonumber\\
&&\lbrack R_{1j},R_{4}]=\frac{1}{2}R_4\delta_{jz}-
\frac{\sqrt{3}}{4}(R_{2x}-i R_{2y})(\delta_{jx}+i\delta_{jy}),\nonumber\\
&&\lbrack R_{2j},R_{3}]=\frac{1}{2}R_3\delta_{jz}-
\frac{\sqrt{3}}{4}(R_{1x}+i R_{1y})(\delta_{jx}+i\delta_{jy}),\nonumber\\
&&\lbrack R_{2j},R_{4}]=-\frac{1}{2}R_4\delta_{jz}+
\frac{\sqrt{3}}{4}(R_{1x}+i R_{1y})(\delta_{jx}-i\delta_{jy}),\nonumber\\
 &&\lbrack R_{3},R_{4}]=\frac{3}{2}(R_{2}^z-R_{1}^z).\label{comu3-s}
\end{eqnarray}
 These operators generate
the algebra $u_3$ in the reduced spin space $\{T1_l,S_l,S_r\}$
specified by the Casimir operator
$${\bf R}^2_1+{\bf R}^2_2+R^2_{3}+R^2_{4}=\frac{3}{2}.$$
Therefore, in this case the TQD possesses $SU(3)$ symmetry. These
$R$ operators may be represented via the familiar Gell-Mann
matrices $\lambda_i$ ($i=1,...,8$) for the $SU(3)$ group,
\begin{eqnarray*}
R^{+}_{1}&=&\frac{1}{2}\left(\lambda_1+i\lambda_2\right), \ \ \ \
\ \
R^{-}_{1}=\frac{1}{2}\left(\lambda_1-i\lambda_2\right), \\
R^z_1&=&\frac{\lambda_3}{2}, \ \ \ \ \ \ \ \ \ \ \ \ \ \ \ \ \ \
R^z_2=\frac{1}{4}(\lambda_3+\sqrt{3}\lambda_8),\\
R^{+}_{2}&=&\frac{1}{2}\left(\lambda_4+i\lambda_5\right), \ \ \ \
\ \
R^{-}_{2}=\frac{1}{2}\left(\lambda_4-i\lambda_5\right), \\
R_{3}&=&\frac{\sqrt{3}}{4}\left(\lambda_6+i\lambda_7\right), \ \ \
 R_{4}=\frac{\sqrt{3}}{4}\left(\lambda_6-i\lambda_7\right).
\end{eqnarray*}

As far as the RG procedure for the "Runge-Lenz" exchange
Hamiltonian (\ref{int7m-s}) the poor-man scaling procedure is
applicable also for the $R$ operators. The scaling equations have
the form,
\begin{eqnarray}
&&\frac{dj_{1}}{d\ln d} = -2j_{3l}^2,\ \ \ \ \ \ \
 \frac{dj_{2}}{d\ln d} =-j_{3r}^2,  \nonumber\\
&&\frac{dj_{3l}}{d\ln d} =-\Big[j_{3l}\Big(j_1+\frac{j_2}{2}\Big)
 -\frac{\sqrt{3}}{4}j_{3r}j_4\Big],  \nonumber\\
&&\frac{dj_{3r}}{d\ln d}=-\frac{j_{3r}(j_1+2j_2
+j_5+2j_6)-\sqrt{3}j_{4}(j_{3l}+\sqrt{2}j_{7})}{4},\nonumber\\
&&\frac{dj_{4}}{d\ln
d}=j_{3r}\Big(\frac{\sqrt{3}}{3}j_{3l}+\frac{\sqrt{2}}{2}j_{7}\Big),
\nonumber\\
&&\frac{dj_{5}}{d\ln d} = -4j_{7}^2,\ \ \ \ \ \ \ \
 \frac{dj_{6}}{d\ln d} =-j_{3r}^2, \nonumber\\
&&\frac{dj_{7}}{d\ln d}
=-\Big[j_{5}j_7+\frac{j_6j_7}{2}-\frac{\sqrt{6}}{8}j_{3r}j_4\Big],
\label{sc7m}
\end{eqnarray}
where $j=\rho_0 J$. We cannot demonstrate analytical solution of
this system, but the numerical solution shows that stable infinite
fixed point exists in this case like in all previous
configurations.

Another type of field induced Kondo effect is realized in the
symmetric case of ${\delta}=E_{S_{g}}-E_{T_{g,u}}<0$. Now the
Zeeman splitting compensates negative $\delta$. Then the two
components of the triplets, namely $E_{T1_{g,u}}$ cross with the
singlet state energy $E_{S_{g}}$, and the symmetry group of the TQD in
magnetic field is $SU(3)$ as in the case considered above.

\subsection{Section summary}
It has been demonstrated
that the loss of rotational invariance in external
magnetic field radically changes the dynamical symmetry of TQD. We
considered here two examples of symmetry reduction, namely
$SO(5)\to SU(3)$ and $P\times SO(4)\times SO(4)\to SU(3)$. In all
cases the Kondo exchange is anisotropic, which, of course,
reflects the axial anisotropy induced by the external field. These
examples as well as the $SO(4)\to SU(2)$ reduction considered
earlier \cite{KA01,KA02} describe the magnetic field induced Kondo
effect owing to the dynamical symmetry of complex quantum dots.
Similar reduction $SO(n)\to SU(n')$ induced by magnetic field may
arise also in more complicated configurations, and in particular
in the parallel geometry. The immense complexity of  scaling
procedure adds nothing new to the general pattern of the field
induced anisotropy of Kondo tunneling, so we confine ourselves
with these two examples .

Although the anisotropic Kondo Hamiltonian was introduced formally
at the early stage of Kondo physics \cite{AYH}, it was rather
difficult to perceive how such Hamiltonian is derivable from the
generic Anderson-type Hamiltonian. It was found that the effective
anisotropy arises in cases where the pseudo-spin degrees of
freedom (like a two-level system) are responsible for anomalous
scattering. Another possibility is the introduction of magnetic
anisotropy in the generic spin Hamiltonian due to spin-orbit
interaction (see Ref. \onlinecite{AZ} for a review of such
models). One should also mention the remarkable possibility of
magnetic field induced anisotropic Kondo effect on a magnetic
impurity in ferromagnetic rare-earth metals with easy plane
magnetic anisotropy.\cite{Sandal} This model is close to our model
from the point of view of effective spin Hamiltonian, but the
sources of anisotropy are different in the two systems. In our
case the interplay between singlet and triplet components of spin
multiplet is an eventual source both of Kondo effect itself and of
its anisotropy in external magnetic field. Previously, the
manifestation of $SU(3)$ symmetry in anisotropic magnetic systems
were established in Ref.\cite{Onufr}. It was shown, in particular,
that this dynamical symmetry predetermines the properties of
collective excitations in anisotropic Heisenberg ferromagnet. In
the presence of single-ion anisotropy the relation between  the
Hubbard operators for $S=1$ and Gell-Mann matrices $\lambda$ were
established. It worth also mentioning in this context the
$SU(4)\supset SO(5)$ algebraic structure of superconducting and
antiferromagnetic coherent states in cuprate High-T$_c$ materials
\cite{foot2}.

\section{Conclusions}
We have analyzed the occurrence of dynamical symmetries in complex
quantum dots. These symmetries emerge when the dot is coupled with
metallic electrodes under the conditions of strong Coulomb
blockade and nearly degenerate low energy spin spectrum.  It can
be achieved either by an application of an external magnetic field
or due to dot-lead tunneling which, as we have seen, results in
level renormalization and the emergence of an additional symmetry.
Although the main focus in this paper is related to the study of
triple quantum dots, the generalization to other quantum dot
structures is indeed straightforward.

Since we were interested in a symmetry aspect of Kondo tunneling
Hamiltonian, we restricted ourselves by derivation of RG flow
equations and solving them for obtaining the Kondo temperature. In
all cases the TQDs possess strong coupling fixed point
characteristic for spin 1/2 and/or spin 1 case. We did not
calculate the tunnel conductance in details, because it reproduces
the main features of Kondo-type zero bias anomalies studied
extensively by many authors (see, e.g.,
\cite{Magn,Glaz,KA01,KA02,KKA02,Pust,Eto02}). The novel feature is
the possibility of changing $T_K$ by scanning the phase diagram of
Fig. \ref{phasediag}. Then the zero bias anomaly follows all
symmetry crossovers induced by experimentally tunable gate
voltages and tunneling rates.

The main message of our work is that symmetry enters the realm of
mesoscopic physics in a rather non-trivial manner. Dynamical
symmetry in this context is not just a geometrical concept but,
rather, intimately related with the physics of strong correlations
and exchange interactions. The relation with other branches of
physics makes it even more attractive. The groups $SO(n)$ play an
important role in Particle Physics as well as in model building
for high temperature superconductivity (especially $SO(5)$). The
role of the group $SU(3)$ in Particle Physics cannot be
overestimated and its role in Nuclear Physics in relation with the
interacting Boson model is well recognized. This paper extends the
role of these Lie groups in Condensed Matter Physics.

\noindent {\bf Acknowledgment}: This work is partially supported
by grants from the Israeli Science Foundations, the United
States Israel Binational Science Foundation and
Deutsche-Israeli-Project (DIP). One of us (T.K) is
deeply indebted to the {\it Clore Scholars Programme}
for generous support.
\appendix
\section{Diagonalization of the dot Hamiltonian}\label{diag}
Here we describe the diagonalization procedure of the Hamiltonian
of the isolated TQDs occupied by four and three electrons. The dot
Hamiltonian has the form,
\begin{eqnarray}
H_{d}&=&\sum_{a=l,r,c}\sum_{\sigma}\epsilon_{a}d^\dagger_{a
\sigma}d_{a\sigma
}+\sum_{a}Q_an_{a\uparrow}n_{a\downarrow}\nonumber\\
 &+&\sum_{a=l,r}(W_{a}d^\dagger_{c
\sigma}d_{a\sigma }+H.c.). \label{H-dot}
\end{eqnarray}
({\bf a}) Four electron occupation:\\
The Hamiltonian
(\ref{H-dot}) can be diagonalized  by using the basis of
four-electron wave functions
\begin{eqnarray}
&&\left| t_a, 1 \right\rangle =
d^{\dagger}_{c\uparrow}d^{\dag}_{a\uparrow}d^{\dag}_{{\bar
a}\uparrow} d^{\dag}_{{\bar a}\downarrow}\vert 0\rangle,\ \ \
\left| t_a, {\bar 1} \right\rangle
=d^{\dagger}_{c\downarrow}d^{\dag}_{a\downarrow}d^{\dag}_{{\bar
a}\uparrow}
d^{\dag}_{{\bar a}\downarrow}\vert 0\rangle,\nonumber \\
&& \left| t_a, 0\right\rangle = \frac{1}{\sqrt{2}}\left (
d^{\dag}_{c\uparrow}d^{\dag}_{a\downarrow}
+d^{\dag}_{c\downarrow}d^{\dag}_{a\uparrow}\right )
d^{\dag}_{{\bar a}\uparrow}d^{\dag}_{{\bar a}\downarrow}\vert 0\rangle
,\nonumber \\
&&\left| s_a \right\rangle =\frac{1}{\sqrt{2}}\left (
d^{\dag}_{c\uparrow}d^{\dag}_{a\downarrow}
-d^{\dag}_{c\downarrow}d^{\dag}_{a\uparrow}\right )
d^{\dag}_{{\bar a}\uparrow}d^{\dag}_{{\bar a}\downarrow}\vert 0\rangle ,
\nonumber \\
&&|{ex}\rangle = d^{\dag}_{l\uparrow}d^{\dag}_{l\downarrow}
d^{\dag}_{r\uparrow}d^{\dag}_{r\downarrow} \vert 0\rangle,
\label{basis-f}
\end{eqnarray}
where $a =l,r$; $\overline l=r$, $\overline r=l$. The Coulomb
interaction quenches the states with two electrons in the central
dot and we do not take them into account. The states
(\ref{basis-f}) form a basis of two triplet and three singlet
states. In this basis, the Hamiltonian (\ref{H-dot}) is decomposed
into triplet and singlet matrices,
\begin{eqnarray}
H_t=\left(
\begin{array}{cc}
\tilde{\varepsilon}_l & 0 \\
0& \tilde{\varepsilon}_r
 \end{array}\right ),
\end{eqnarray}
and
\begin{eqnarray}
H_s= \left(
\begin{array}{ccc}
\tilde{\varepsilon}_l & 0 & \sqrt{2}W_l \\
0 & \tilde{\varepsilon}_r & \sqrt{2}W_r \\
\sqrt{2}W_l & \sqrt{2}W_r & \tilde{\varepsilon}_{ex} \\
\end{array}\right ),
\end{eqnarray}
where $\tilde{\varepsilon}_l={\epsilon}_c
+{\epsilon}_l+2{\epsilon}_r+Q_r$,
$\tilde{\varepsilon}_r={\epsilon}_c
+{\epsilon}_r+2{\epsilon}_l+Q_l$, and $\tilde{\varepsilon}_{ex}
=2{\epsilon}_l+2{\epsilon}_r+Q_l+Q_r.$ We are interested in the
limit $\beta_a\ll 1$ ($\beta_a$ are defined by Eq.(15)). So the
secular matrix may be diagonalized in lowest order of perturbation
theory in $\beta_a$.
 The eigenfunctions corresponding to the
energy levels (22) are
\begin{eqnarray}
&&\vert S_a\rangle = \sqrt{1-2\beta_a^2}\left\vert
s_a\right\rangle
-\sqrt{2}\beta_a|{ex}\rangle, \nonumber \\
&&\vert T_a\rangle = \left\vert t_a\right\rangle,
\label{eg-fun4}\\
&&\vert {Ex}\rangle = \sqrt{1-2\beta_l^2-2\beta_r^2}\left\vert
{ex}\right\rangle+ \sqrt{2}\beta_l|s_l\rangle
+\sqrt{2}\beta_r|s_r\rangle.\nonumber
\end{eqnarray}

In completely symmetric case, $\varepsilon_l=\varepsilon_r\equiv
\varepsilon,$ $Q_l=Q_r\equiv Q,$ $W_l=W_r\equiv W$, the
eigenfunctions corresponding to the energies (\ref{degen}) are
\begin{eqnarray}
\vert S_+\rangle &=& \sqrt{1-4\beta^2}\frac{\left\vert
s_l\right\rangle+\left\vert s_r\right\rangle}{\sqrt{2}}
-2\beta|{ex}\rangle, \nonumber \\
\vert S_-\rangle &=& \frac{\left\vert s_l\right\rangle-\left\vert
s_r\right\rangle}{\sqrt{2}},
\label{degen-eg-fun}\\
\vert T_{\pm}\rangle &=& \frac{\left\vert t_l\right\rangle\pm
\left\vert
t_r\right\rangle}{\sqrt{2}},\nonumber\\
\vert {Ex}\rangle &=& \sqrt{1-4\beta}\left\vert {ex}\right\rangle+
\sqrt{2}\beta(|s_l\rangle +|s_r\rangle),\nonumber
\end{eqnarray}
where $\beta=W/(\varepsilon+Q-\epsilon_c)$.\\
({\bf b}) Three electron occupation:\\
In this case the Hamiltonian (\ref{H-dot}) can be diagonalized by
using the basis of three--electron wave functions
\begin{eqnarray}
&&| b,\sigma\rangle =\frac{(
[d^{+}_{c\uparrow}d^{+}_{l\downarrow}-d^{+}_{c\downarrow}
d^{+}_{l\uparrow}] d^{+}_{r\sigma}+
[d^{+}_{c\downarrow}d^{+}_{r\uparrow}-d^{+}_{c\uparrow}
d^{+}_{r\downarrow}] d^{+}_{l\sigma} ) \vert 0\rangle}{\sqrt{6}}
,\nonumber\\
&&\left| b_c, \sigma\right\rangle =-\frac{1}{\sqrt{2}}\left (
d^{+}_{l\uparrow}d^{+}_{r\downarrow}-d^{+}_{l\downarrow}
d^{+}_{r\uparrow}\right ) d^{+}_{c\sigma}\vert 0\rangle ,\nonumber \\
&&\left| q, {\pm\frac{3}{2}}\right\rangle
=d^{+}_{c\pm}d^{+}_{r\pm}d^{+}_{l\pm}\vert 0\rangle ,
 \nonumber\\
&&\left | q, {\pm\frac{1}{2}}\right\rangle =\frac{(
d^{+}_{c\pm}d^{+}_{r\pm}d^{+}_{l\mp}+
d^{+}_{c\pm}d^{+}_{r\mp}d^{+}_{l\pm}+
d^{+}_{c\mp}d^{+}_{r\pm}d^{+}_{l\pm})\vert
0\rangle }{\sqrt{3}}, \nonumber\\
&&|b_{lc},\sigma\rangle = d^{+}_{l\uparrow}d^{+}_{l\downarrow}
d^{+}_{c\sigma}\vert 0\rangle,\ \ \ \ \ \;\;\;  |b_{rc},
\sigma\rangle = d^{+}_{r\uparrow}d^{+}_{r\downarrow}
d^{+}_{c\sigma} \vert 0\rangle,\nonumber\\
&&|b_l, \sigma\rangle = d^{+}_{r\uparrow}d^{+}_{r\downarrow}
d^{+}_{l\sigma}\vert 0\rangle,\ \ \ \ \ \ \ \ |b_r, \sigma\rangle
= d^{+}_{l\uparrow}d^{+}_{l\downarrow} d^{+}_{r\sigma} \vert
0\rangle,\label{basis}
\end{eqnarray}
where $\sigma=\uparrow,\downarrow$. The three-electron states
$|\Lambda\rangle$ of the TQD are classified as a ground state
doublet $|B_1\rangle$, low-lying doublet $|B_2\rangle$ and quartet
$|Q\rangle$ excitations, and four charge-transfer excitonic
doublets $B_{ac}$ and $B_a$ ($a=l,r$). In the framework of second
order perturbation theory with respect to $\beta_{a}$ (15), the
 energy levels $E_\Lambda$ are
\begin{eqnarray}
&&E_{B_1} = {\epsilon}_c +{\epsilon}_l +{\epsilon}_r-
\frac{3}{2}\left[W_l \beta_l+W_r \beta_r\right], \nonumber \\
&&E_{B_2} = {\epsilon}_c +{\epsilon}_l+{\epsilon}_r-
\frac{1}{2}\left[W_l \beta_l+W_r \beta_r\right],
\nonumber \\
&&E_{Q} = {\epsilon}_c +{\epsilon}_l +\epsilon_r,\nonumber\\
&&E_{B_{ac}}={\epsilon}_c+2{\epsilon}_a+Q_a-W_{\bar a}\beta_{\bar a},\nonumber\\
&&E_{B_a} ={\epsilon}_a +2{\epsilon}_{\bar a}+Q_{\bar
a}+W_{a}\beta_{a}+2W_{\bar a}\beta_{\bar a}. \label{energy3}
\end{eqnarray}

 The eigenfunctions corresponding to the
energy levels (\ref{energy3}) are the following combinations,
\begin{eqnarray}
&&\vert B_1,\sigma\rangle = \gamma_1\vert b_1,\sigma\rangle
-\frac{\sqrt{6}}{2}\beta_{l}|b_{r},\sigma\rangle +
\frac{\sqrt{6}}{2}\beta_{r}|b_{l},\sigma\rangle , \nonumber\\
&&\vert B_2,\sigma\rangle = \gamma_2\vert b_2,\sigma\rangle
-\frac{\sqrt{2}}{2}\beta_{l}|b_{r},\sigma\rangle -
\frac{\sqrt{2}}{2}\beta_{r}|b_{l},\sigma\rangle , \nonumber\\
&&\vert Q,s_z\rangle =\left\vert q,s_z\right\rangle,\ \ \ \ \ \
s_z=\pm \frac{3}{2},\pm \frac{1}{2},\label{eg-func}\\
&&\vert B_{ac},\sigma\rangle =\sqrt{1-\beta_{\bar a}^2}\vert
b_{ac},\sigma\rangle
-\beta_{\bar a}|b_{\bar a},\sigma\rangle,\nonumber\\
&&\vert B_{r},\sigma\rangle = \sqrt{1-2\beta_{l}^2-\beta_r^2}\vert
b_r,\sigma\rangle+ \beta_{r}|b_{lc},\sigma\rangle\nonumber\\
&&\ \ \ \ \ \ \ \ \ \
+\frac{\sqrt{2}}{2}\beta_{l}\left(\sqrt{3}|b_1,\sigma\rangle
+|b_2,\sigma\rangle\right), \nonumber\\
&&\vert B_l,\sigma\rangle = \sqrt{1-2\beta_r^2-\beta_l^2}\vert
b_l,\sigma\rangle +\beta_{l}|b_{r},
{c},\sigma\rangle\nonumber\\
&&\ \ \ \ \ \ \ \ \ \;
-\frac{\sqrt{2}}{2}\beta_{r}\left(\sqrt{3}|b_1,\sigma\rangle
-|b_2,\sigma\rangle\right) ,\nonumber
\end{eqnarray}
where $\gamma_1$ and $\gamma_2$ are determined by Eq.(\ref{gam}).

\section{Rotations in the source-drain and left-right
space}\label{GR-trans}

In the generic case, the transformation which diagonalizes the
tunneling Hamiltonian (\ref{hyb}) has the form
\begin{equation}
\left(
\begin{array}{cc} c_{lek\sigma }\\ c_{lok\sigma }\\ c_{rek\sigma }\\ c_{rok\sigma }\\
\end{array} \right) =
\left(
\begin{array}{rrrr} u_l & v_l & 0 & 0 \\
                 -v_l & u_l & 0 & 0 \\
                 0 & 0 & u_r & v_r \\
                 0 & 0 & -v_r & u_r
\end{array} \right)
\left(
\begin{array}{cc} c_{lsk\sigma }\\ c_{ldk\sigma }\\ c_{rsk\sigma }\\ c_{rdk\sigma }
\end{array} \right)
\end{equation}
with $u_a=V_{as}/V_{a}$, $v_a=V_{ad}/V_{a}$; $V_{a}^2=|V_{as}|^2 +
|V_{ad}|^2$ $(a=l,r).$ In a symmetric case $V_{as}=V_{ad}=V$, this
transformation simplifies  to
\begin{eqnarray}
&&c_{aek\sigma }=2^{-1/2}\left(c_{ask\sigma }+c_{adk\sigma }\right),\nonumber \\
&&c_{aok\sigma }=2^{-1/2}\left(-c_{ask\sigma }+c_{adk\sigma
}\right),
\end{eqnarray}
and only the even $(e)$ combination survives in the tunneling
Hamiltonian
\begin{equation}
H_{tun}=V\sum_{ak\sigma }(c_{aek\sigma }^\dag d_{a\sigma} + H.c.).
\end{equation}
So the odd combination  $(o)$ may be omitted.

If, moreover, the whole system "TQD plus leads" possesses $l-r$
symmetry, $\varepsilon_{l}=\varepsilon_r$, the second rotation in
$l-r$-space
\begin{equation}
\left(
\begin{array}{ll} c_{gk\sigma }\\ c_{uk\sigma }\\ d_{g\sigma }\\ d_{u\sigma }\\
\end{array} \right) =\frac{1}{\sqrt 2}
\left(
\begin{array}{rrrr} 1 & 1 & 0 & 0 \\
                 -1 & 1 & 0 & 0 \\
                 0 & 0 & 1 & 1 \\
                 0 & 0 & -1 & 1
\end{array} \right)
\left(
\begin{array}{ll} c_{lek\sigma }\\ c_{rek\sigma }\\ d_{l\sigma }\\ d_{r\sigma }
\end{array} \right)
\end{equation}
transforms $H_{lead}+H_{tun}$ into
\begin{eqnarray}
H_{lead}+H_{tun}=\sum_{\eta k\sigma}\Big[ \epsilon_{k\eta}n_{\eta
k\sigma} + V(c_{\eta k\sigma }^\dag d_{\eta\sigma} + H.c.)\Big]
\end{eqnarray}
with $\epsilon_{kg}=\epsilon_{k}-t_{lr}$,
$\epsilon_{ku}=\epsilon_{k}+t_{lr}$.

\section{Effective spin Hamiltonian}\label{H-spin}
The spin Hamiltonian of the TQD with $N=4$ occupation in series
geometry (Fig.\ref{TQD-s}) is derived below. The system is
described by the Hamiltonian (\ref{Hser}). The Schrieffer-Wolff
transformation \cite{SW} for the configuration of four electron
states of the TQD projects out three electron states $\vert
\lambda\rangle$ and maps the Hamiltonian (\ref{Hser}) onto an
effective spin Hamiltonian $\widetilde{H}$ acting in a subspace of
four-electron configurations $\langle\Lambda|\ldots
|\Lambda^{\prime}\rangle$,
\begin{eqnarray}\label{SW-tr}
\widetilde{H}= e^{i{\cal S}}H e^{{-i\cal S}}
= H + \sum_m \frac{(i)^m}{m!} [%
{\cal S},[{\cal S}...[{\cal S},H]]...],
\end{eqnarray}
where
\begin{equation}
{\cal S}=-i\sum_{\Lambda\lambda}\sum_{\langle k\rangle\sigma,a}
\frac{V_{a\sigma}^{\Lambda\lambda}}
{\bar{E}_{\Lambda\lambda}-\epsilon_{ka}}
X^{\Lambda\lambda}c_{ak\sigma}+H.c.
\end{equation}
Here $\langle k\rangle$ stands for the electron or hole states
whose energies are secluded
within a layer $\pm \bar{D}$ around the Fermi level. $\bar{E}%
_{\Lambda\lambda}=E_\Lambda(\bar{D})-E_\lambda(\bar{D})$ and the
notation $a=l,r$ is used.
 The effective Hamiltonian with three--electron states
$|\lambda\rangle$ frozen out can be obtained by retaining the
terms to order $O(|V|^2)$ on the right-hand side of
Eq.(\ref{SW-tr}). It has the following form,
\begin{eqnarray}
\widetilde{H}  &=&  \sum_\Lambda \bar{E}_\Lambda
X^{\Lambda\Lambda} +\sum_{\langle k\rangle \sigma,a}\epsilon_{ka}
c^{+}_{ak\sigma }c_{ak\sigma}
\nonumber \\
& - &
\sum_{\Lambda\Lambda^{\prime}\lambda}\sum_{kk^{\prime}\sigma\sigma^{\prime}}
\sum_{a=l,r} J^{\Lambda\Lambda^{\prime}}_{kk^{\prime}a}
X^{\Lambda\Lambda^{\prime}}c^{+}_{ak\sigma}c_{ak^{\prime}\sigma^{\prime}}
\label{Ham-eff} \\
& - &
\sum_{\Lambda\Lambda^{\prime}\lambda}\sum_{kk^{\prime}\sigma\sigma^{\prime}}
( J^{\Lambda\Lambda^{\prime}}_{kk^{\prime}lr}
X^{\Lambda\Lambda^{\prime}}c^{+}_{rk\sigma}
c_{lk^{\prime}\sigma^{\prime}}+H.c.),\nonumber
\end{eqnarray}
where
\begin{eqnarray}
J^{\Lambda\Lambda^{\prime}}_{kk^{\prime}a} = \frac
{(V_{a\sigma}^{\lambda\Lambda})^*
V_{a\sigma^{\prime}}^{\lambda\Lambda^{\prime}}}{2} \left(
\frac{1}{\bar{E}_{\Lambda\lambda}-\epsilon_{ka}}+
\frac{1}{\bar{E}_{\Lambda^{\prime}\lambda}-\epsilon_{k^{\prime}a}}
\right),\nonumber \\
J^{\Lambda\Lambda^{\prime}}_{kk^{\prime}lr} = \frac
{(V_{l\sigma}^{\lambda\Lambda})^*
V_{r\sigma^{\prime}}^{\lambda\Lambda^{\prime}}} {2} \left(
\frac{1}{\bar{E}_{\Lambda\lambda}-\epsilon_{kl}}+
\frac{1}{\bar{E}_{\Lambda^{\prime}\lambda}-\epsilon_{k^{\prime}r}}
\right).
\end{eqnarray}
The constraint $\sum_{\Lambda}X^{\Lambda%
\Lambda}=1 $ is valid. Unlike the conventional case 
of doublet spin $1/2$ we have here an octet
$\Lambda=\{\Lambda_l,\Lambda_r\}=\{S_l,T_l,S_r,T_r\}$, and the SW
transformation {\it intermixes all these states}. The effective
spin Hamiltonian (\ref{Ham-eff}) to order $O(|V|^2)$ acquires the
form of Eq.(\ref{ex-H}).

\section{$\bf SO(7)$ symmetry}\label{algebra-o7}

The operators ${\bf S}_l$, ${\bf S}_r$, ${\bf R}_l$, $\tilde{\bf
R}_1$, $\tilde{\bf R}_2$, $\tilde{\bf R}_3$ and ${A_i}$
$(i=1,2,3)$ (see Eqs.(33),(\ref{so7}),(\ref{A})) obey the
commutation relations of the $o_7$ Lie algebra,
\begin{eqnarray*}
&&\lbrack S_{aj},S_{a'k}]=ie_{jkm}\delta_{aa'}S_{am},\ \ \;\;
[R_{lj},R_{lk}]=ie_{jkm}S_{lm},\nonumber\\
&&[R_{lj},S_{lk}]=ie_{jkm}R_{lm},\ \ \ \ \ \ \ \ \ \
[R_{lj},S_{rk}]=[\tilde{R}_{3j},S_{lk}]=0,\nonumber\\
&&\lbrack \tilde{R}_{3j},\tilde{R}_{3k}] = ie_{jkm}S_{rm},\ \ \ \
\ \ \ \ \
[\tilde{R}_{3j},S_{rk}]=ie_{jkm}\tilde{R}_{3m},\nonumber\\
&&[\tilde{R}_{1j},\tilde{R}_{1k}]=ie_{jkm}S_{rm}(1-{\delta}_{jz})
(1-{\delta}_{kz})\nonumber\\
&&\ \ \  \ \ \ \ \ \ \ \ \ \;
+\frac{i}{2}e_{jkm}S_{lm}({\delta}_{jz}+{\delta}_{kz})-
\frac{1}{2}(S_{lj}{\delta}_{kz}-S_{lk}{\delta}_{jz}),\nonumber\\
&&[\tilde{R}_{2j},\tilde{R}_{2k}]=ie_{jkm}S_{lm}(1-{\delta}_{jz})
(1-{\delta}_{kz})\nonumber\\
&&\ \ \  \ \ \ \ \ \ \ \ \ \;
+\frac{i}{2}e_{jkm}S_{rm}({\delta}_{jz}+{\delta}_{kz})-
\frac{1}{2}(S_{rj}{\delta}_{kz}-S_{rk}{\delta}_{jz}),\nonumber\\
&&[\tilde{R}_{1j},\tilde{R}_{2k}]=\frac{i}{2}e_{jkm}(S_{rm}\delta_{jz}
+S_{lm}{\delta}_{kz})\nonumber\\
&&\ \ \  \ \ \ \ \ \ \ \ \ \
+\frac{1}{2}\left[S_{lj}{\delta}_{kz}-S_{rk}{\delta}_{jz}+
(S_{lz}-S_{rz})\delta_{jz}\delta_{kz}\right],\nonumber\\
&&\lbrack \tilde{R}_{3j},\tilde{R}_{1k}] =
ie_{jkm}R_{lm}(1-{\delta}_{jz}-\frac{{\delta}_{kz}}{2})\nonumber\\
&&\ \ \  \ \ \ \ \ \ \ \ \ \ \
-\frac{{\delta}_{kz}}{2}(1-{\delta}_{jz})R_{lj},\nonumber\\
&&\lbrack \tilde{R}_{3j},\tilde{R}_{2k}] =
ie_{jkm}R_{lm}({\delta}_{jz}+\frac{{\delta}_{kz}}{2})+
\frac{{\delta}_{kz}}{2}(1-{\delta}_{jz})R_{lj},\nonumber\\
&&[\tilde{R}_{1j},R_{lk}]=ie_{jkm}\tilde{R}_{3m}({\delta}_{kz}+
\frac{{\delta}_{jz}}{2})-
\frac{{\delta}_{jz}}{2}(1-{\delta}_{kz})\tilde{R}_{3k},\nonumber\\
&&[\tilde{R}_{2j},R_{lk}]=ie_{jkm}\tilde{R}_{3m}(1-{\delta}_{kz}-
\frac{{\delta}_{jz}}{2})\nonumber\\
&&\ \ \  \ \ \ \ \ \ \ \ \ \
+\frac{{\delta}_{jz}}{2}(1-{\delta}_{kz})\tilde{R}_{3k},
\end{eqnarray*}
\vspace{-6mm}
\begin{eqnarray*}
&&[A_1,S_{lj}]=iA_2{\delta}_{jz}+\frac{i\sqrt{2}}{2}
(\tilde{R}_{1x}\delta_{jx}-\tilde{R}_{1y}\delta_{jy}),\nonumber\\
&&[A_2,S_{lj}]=-iA_1{\delta}_{jz}-\frac{i\sqrt{2}}{2}
(\tilde{R}_{1y}\delta_{jx}+\tilde{R}_{1x}\delta_{jy}),\nonumber\\
&&[A_1,S_{rj}]=iA_2{\delta}_{jz}-\frac{i\sqrt{2}}{2}
(\tilde{R}_{2x}\delta_{jx}-\tilde{R}_{2y}\delta_{jy}),\nonumber\\
&&[A_2,S_{rj}]=-iA_1{\delta}_{jz}+\frac{i\sqrt{2}}{2}
(\tilde{R}_{2y}\delta_{jx}+\tilde{R}_{2x}\delta_{jy}),\nonumber\\
&&[A_3,S_{lj}]=-i\tilde{R}_{2j}(1-\delta_{jz}),\ \ \
[A_3,S_{rj}]=i\tilde{R}_{1j}(1-\delta_{jz}),
\end{eqnarray*}
\vspace{-6mm}
\begin{eqnarray}
&&[A_{1},R_{lj}]=-\frac{i\sqrt{2}}{2}(\tilde{R}_{3x}\delta_{jx}-
\tilde{R}_{3y}{\delta}_{jy}),\nonumber\\
&&[A_{2},R_{lj}]=\frac{i\sqrt{2}}{2}
(\tilde{R}_{3y}\delta_{jx}+\tilde{R}_{3x}{\delta}_{jy}),\nonumber\\
&&[A_{1},\tilde{R}_{3j}]=\frac{i\sqrt{2}}{2}
(R_{lx}\delta_{jx}-R_{ly}{\delta}_{jy}),\nonumber\\
&&[A_{2},\tilde{R}_{3j}]=-\frac{i\sqrt{2}}{2}(R_{ly}\delta_{jx}+
R_{lx}{\delta}_{jy}),\nonumber\\
&&[A_{3},R_{lj}]=-i\tilde{R}_{3z}{\delta}_{jz},\;\;\;\;\;\;\;\;
[A_{3},\tilde{R}_{3j}]=iR_{lz}{\delta}_{jz},\nonumber\\
&&[A_{1},\tilde{R}_{1j}]=-\frac{i\sqrt{2}}{2}
(S_{lx}\delta_{jx}-S_{ly}{\delta}_{jy}),\nonumber\\
&&[A_{2},\tilde{R}_{1j}]=\frac{i\sqrt{2}}{2}
(S_{ly}\delta_{jx}+S_{lx}{\delta}_{jy}),\nonumber\\
&&[A_{3},\tilde{R}_{1j}]=-i(S_{rx}\delta_{jx}+S_{ry}{\delta}_{jy}),\nonumber\\
&&[A_{1},\tilde{R}_{2j}]=\frac{i\sqrt{2}}{2}(S_{rx}\delta_{jx}-
S_{ry}{\delta}_{jy}),\nonumber\\
&&[A_{2},\tilde{R}_{2j}]=-\frac{i\sqrt{2}}{2}
(S_{ry}\delta_{jx}+S_{rx}{\delta}_{jy}),\nonumber\\
&&[A_{3},\tilde{R}_{2j}]=i(S_{lx}\delta_{jx}+S_{ly}{\delta}_{jy}),\nonumber\\
&&[A_{1},A_{2}]=-i(S_{lz}+S_{rz}),\ \ \ \ \
[A_{1},A_{3}]=[A_{2},A_{3}]=0,\nonumber\\
&&\lbrack S_{aj},\tilde{R}_{\mu k}]=\tau_{jkm}^{a\mu\nu}
\tilde{R}_{\mu m}+\alpha_{jk}^{a\mu n}A_n,\nonumber\\
&&[\tilde{R}_{3j},R_{lk}]=\beta_{jkm}^{\mu}\tilde{R}_{\mu
m}+\tilde \alpha_{jk}^{n}A_n.
\end{eqnarray}
Here $j,k,m$ are Cartesian indices, $a=l,r$; $\mu,\nu =1,2$;
$n=1,2,3$; $\tau_{jkm}^{a\mu\nu}$, $\alpha_{jk}^{a\mu n}$, $\tilde
\alpha_{jk}^{n}$ and $\beta_{jkm}^{\mu}$ are the structural
constants, $\tau_{jkm}^{l\mu\nu}=\tau_{jkm}^{r{\bar \mu}{\bar
\nu}}$, $\alpha_{jk}^{l \mu n}=-\alpha_{jk}^{r {\bar \mu} n}$
(${\bar 1}=2$, ${\bar 2}=1$). Their non-zero components are:
\begin{eqnarray*}
\tau_{xxz}^{l11}&=&\tau_{xzx}^{l11}=\tau_{yyz}^{l11}=\tau_{yzy}^{l11}=\frac{1}{2
},\\
\tau_{xzx}^{l21}&=&\tau_{yzy}^{l21}=\tau_{xxz}^{l12}=\tau_{yyz}^{l12}=-\frac{1}{
2},\\
\tau_{xyz}^{l11}&=&\tau_{xyz}^{l12}=\tau_{yzx}^{l21}=\frac{i}{2},\\
\tau_{xzy}^{l11}&=&\tau_{yxz}^{l11}=\tau_{yxz}^{l12}=\tau_{xzy}^{l21}=-\frac{i}{
2},\\
\tau_{zzz}^{l11}&=&1,\ \tau_{zzz}^{l22}=-1,\ \tau_{zxy}^{l22}=i,\
\tau_{zyx}^{l22}=-i,
\end{eqnarray*}
\vspace{-7mm}
\begin{eqnarray*}
\alpha_{xy}^{l11}&=&\alpha_{yx}^{l11}=\frac{\sqrt{2}}{2},\ \ \
\alpha_{xy}^{l12}=\alpha_{yx}^{l12}=-\frac{\sqrt{2}}{2},\\
\alpha_{xx}^{l11}&=&\alpha_{xx}^{l12}=\frac{i\sqrt{2}}{2},\ \
\alpha_{yy}^{l11}=\alpha_{yy}^{l12}=-\frac{i\sqrt{2}}{2},\\
\alpha_{xx}^{l23}&=&\alpha_{yy}^{l23}=-i\sqrt{2},
\end{eqnarray*}
\vspace{-7mm}
\begin{eqnarray*}
\beta_{xxz}^{1}&=&\beta_{yyz}^{2}=-\frac{1}{2},\ \ \
\beta_{xxz}^{2}=\beta_{yyz}^{1}=\frac{1}{2},\\
\beta_{xyz}^{1}&=&\beta_{xyz}^{2}=\frac{i}{2},\ \ \ \ \
\beta_{yxz}^{1}=\beta_{yxz}^{2}=-\frac{i}{2},\\
\beta_{xzy}^{1}&=&\beta_{zyx}^{2}=-i,\ \ \ \
\beta_{yzx}^{1}=\beta_{zxy}^{2}=i,
\end{eqnarray*}
\vspace{-7mm}
\begin{eqnarray*}
\tilde\alpha_{xx}^{1}&=&\tilde\alpha_{zz}^{3}=i\sqrt{2},\ \ \
\tilde\alpha_{xy}^{2}=\tilde\alpha_{yx}^{2}=\tilde\alpha_{yy}^{1}=-i\sqrt{2}.
\end{eqnarray*}

The following relations hold,
\begin{eqnarray}
&&{\bf S}_a\cdot {\bf R}_l={\bf S}_a\cdot \tilde{\bf R}_3=0,\ \ \
A_1A_3=A_2A_3=0,\nonumber\\
&&{\bf S}^{2}_a=2X^{\mu_a\mu_a},\;\tilde{\bf R}_1\cdot \tilde{\bf
R}^{\dag}_1+\tilde{\bf R}_2\cdot \tilde{\bf R}^{\dag}_2=
2\sum_{a=l,r}X^{\mu_a\mu_a},\nonumber\\
&&{\bf R}^{2}_l=X^{\mu_l\mu_l}+3X^{S_l
S_l},\;\;\;\;\tilde{\bf R}^{2}_3=X^{\mu_r\mu_r}+3X^{S_l S_l}, \nonumber\\
&&A_1^2+A_2^2+A_3^2=X^{\mu_l\mu_l}+X^{\mu_r\mu_r}.
\end{eqnarray}
Therefore, the vector operators ${\bf S}_l$, ${\bf S}_r$, ${\bf
R}_l$, $\tilde{\bf R}_i$ and scalar operators $A_i$ $(i=1,2,3)$
generate the algebra $o_{7}$ in a representation specified by the
Casimir operator
\begin{eqnarray}
{\bf S}_l^{2}+{\bf S}_r^{2}+{\bf R}^{2}_l+\sum_{i=1}^{2}\tilde{\bf
R}_i\cdot \tilde{\bf R}^{\dag}_i+\tilde{\bf
R}^{2}_3+\sum_{i=1}^3A_i^2=6.
\end{eqnarray}


\section{Young tableaux corresponding to various symmetries}\label{tab}

\begin{figure}[htb]
\centering
\includegraphics[width=30mm,height=28mm,angle=0]{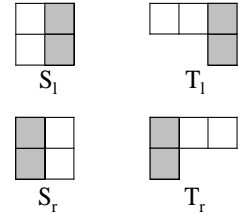}
\caption{Young tableaux corresponding to the singlet ($S_a$) and
triplet ($T_a$) four electron states of the TQD. The grey column
denote two electrons in the same dot (right or
left).}\label{scheme}
\end{figure}

A TQD with "passive" central dot and "active" side dots reminds an
artificial atom with inner core and external valence shell. The
many-electron wave functions in this nanoobject may be symmetrized
in various ways, so that each spin state of $N$ electrons in the
TQD is characterized by its own symmetrization scheme. One may
illustrate these schemes by means of the conventional graphic
presentation of the permutation symmetry of multi-electron
system by Young tableau
 \cite{ED}. For instance, triplet state of two electrons
which is symmetric with respect to the electron permutation is
labeled by a row of two squares, whereas the singlet one which is
antisymmetric with respect to the permutation is labeled by a
column of two squares.
 Having this in mind we can represent the singlet and triplet four
 electron states of the TQD (\ref{eg-fun4}) by the four tableaux shown in
 Fig.\ref{scheme}. The tableaux $S_l$ ($S_r$) and $T_l$ ($T_r$) correspond to
the singlet and triplet states in which the right (left) dot
contains two electrons (grey column in Fig.\ref{scheme}) whereas
electrons in the left (right) and central dots form singlet and
triplet, respectively.

\begin{figure}[htb]
\centering
\includegraphics[width=85mm,height=68mm,angle=0]{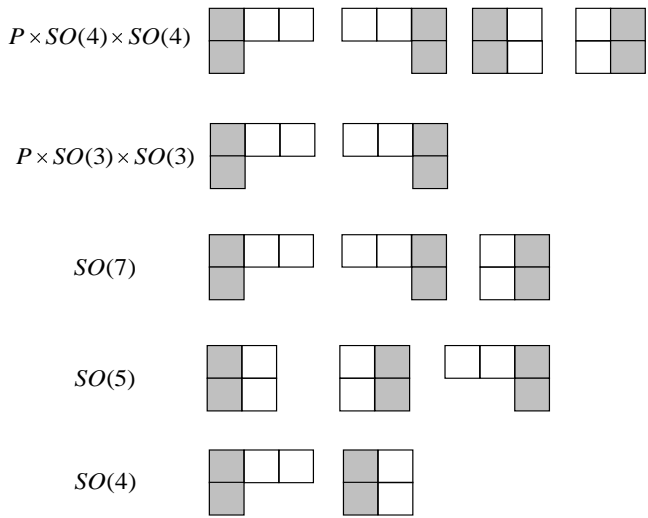}
\caption{Young tableaux corresponding to $SO(n)$
symmetries.}\label{sch-sym}
\end{figure}
The Young tableaux corresponding to various $SO(n)$ symmetries
discussed in Sec.III can be obtained by combining the appropriate
tableaux (Fig. \ref{sch-sym}). The highest possible symmetry
$P\times SO(4)\times SO(4)$ is represented by four tableaux $T_l$,
$T_r$, $S_l$ and $S_r$ since all singlet and triplet states are
degenerate in this case. The symmetry $P\times SO(3)\times SO(3)$
occurs when two triplets $T_l$ and $T_r$ are close in energy and
these are represented by the couple of Young tableaux in the
second line. Following this procedure, the $SO(7)$ symmetry can be
described in terms of two triplets $T_l$, $T_r$ diagrams and one
singlet $S_l$ diagram. Moreover, $SO(5)$ symmetry is represented
by two singlet $S_l$, $S_r$ diagrams and one triplet $T_l$ diagram
and, finally, one triplet and one singlet tableaux correspond to
the $SO(4)$ symmetry.

\end{document}